\documentstyle[twoside,fleqn,epsfig]{article}
\graphicspath{{./fig/}}
\newcommand{\be}{\begin{equation}}
\newcommand{\ee}{\end{equation}}
\newcommand{\bea}{\begin{eqnarray}}
\newcommand{\eea}{\end{eqnarray}}

\newcommand{\AmS}{{\protect\the\textfont2
  A\kern-.1667em\lower.5ex\h\citebox{M}\kern-.125emS}}
\newcommand{\lsim}{\mathrel{\mathop{\kern 0pt \rlap
  {\raise.2ex\hbox{$<$}}}
  \lower.9ex\hbox{\kern-.190em $\sim$}}}
\newcommand{\gsim}{\mathrel{\mathop{\kern 0pt \rlap
  {\raise.2ex\hbox{$>$}}}
  \lower.9ex\hbox{\kern-.190em $\sim$}}}
\def\Journal#1#2#3#4{{#1} {#2} (#3) #4 }

\def\APP{\em Astrop. Phys.}
\def\NPA{{\em Nucl. Phys.} A}
\def\NCAA{{\em Il Nuovo Cimento} A}
\def\NCC{{\em Il Nuovo Cimento} C}

\def\NIMA{{\em Nucl. Instr. \& Methods} A}
\def\NJP{\em New Journal of Physics}

\def\NPB{{\em Nucl. Phys.} B}

\def\PLB{{\em Phys. Lett.} B}

\def\PRL{\em Phys. Rev. Lett.}
\def\PREV{\em Phys. Rev.}

\def\PRD{{\em Phys. Rev.} D}
\def\PRC{{\em Phys. Rev.} C}

\normalsize
\begin{document}

\baselineskip=0.65cm
\vspace*{0.5cm}

\begin{center}
\Large \bf Dark Matter particles in the galactic halo: results and implications from DAMA/NaI\\
\rm
\end{center}

\vspace{0.5cm}
\normalsize

\noindent \rm
R.\,Bernabei,~P.\,Belli,~F.\,Cappella,~R.\,Cerulli,~F.\,Montecchia\footnote{also:
Universita' "Campus Bio-Medico" di Roma, 00155, Rome, Italy},~F.\,Nozzoli

\noindent {\it Dip. di Fisica, Universita' di Roma "Tor Vergata"
and INFN, sez. Roma2, I-00133 Rome, Italy}

\vspace{3mm}   

\noindent \rm A.\,Incicchitti,~D.\,Prosperi

\noindent {\it Dip. di Fisica, Universita' di Roma "La Sapienza"
and INFN, sez. Roma, I-00185 Rome, Italy}

\vspace{3mm}

\noindent \rm C.J.\,Dai,~H.H.\,Kuang,~J.M.\,Ma,~Z.P.\,Ye\footnote{also:
University of Zhao Qing, Guang Dong, China}

\noindent {\it IHEP, Chinese Academy, P.O. Box 918/3, Beijing 100039, China}

\vspace{1cm}

\normalsize

\begin{abstract}
The DAMA/NaI experiment ($\simeq$ 100 kg highly radiopure NaI(Tl)) 
was proposed, designed and realised to effectively investigate in a model independent way
the presence of a Dark Matter particle component in the galactic 
halo by exploiting the annual modulation signature.
With a total exposure of 107731 kg $\cdot$ day, collected 
over seven annual cycles deep underground at the Gran Sasso National
Laboratory of the I.N.F.N., it has pointed out -- 
at 6.3 $\sigma$ C.L. -- an effect which satisfies all the peculiarities of the signature and
neither systematic effects nor side reactions able to mimic the signature
were found. Moreover, several (but still few with respect to the possibilities) corollary model dependent 
quests for the candidate particle have been carried out.
In this paper the obtained results are summarized and some 
perspectives are discussed at some extent.
\end{abstract}



\section{Introduction}

The problem of the existence of Dark Matter in our Universe dates back to the 
astrophysical observations at the beginning of past century \cite{Zwi,Smi}, but
the presence of Dark Matter in our Universe 
has been definitively accepted by the scientific community only about 
40 years later, when two groups performed systematic measurements 
of the rotational velocities of celestial bodies in spiral galaxies \cite{Spi}.
After the '70 many other observations have further confirmed the presence of Dark Matter
in the Universe and, at present, the measurements are mainly devoted to 
the investigation of the quantity, of the distribution (from the cosmological scale down to the galactic one) 
and of the nature of the Dark Matter in the Universe.
Recent measurements of the CMB temperature anisotropy
by WMAP \cite{Wma}, analysed in the framework of the 
{\it{Big Bang}} cosmological scenario, support
a density of the Universe: $\Omega$ = 1, further crediting that most of the Universe is dark.
Recently, it has been suggested from 
observations on the supernovae Ia at high red-shift as standard candles that 
about 73\% of $\Omega$ might 
be in form of a {\it dark energy} \cite{RiPe};
the argument is still under investigation
and presents some problems on possible theoretical interpretations.
However, even in this scenario -- where the matter density in the Universe would be $\Omega_m \sim 0.3$ \cite{Wma,BoWm} --
large space for Dark Matter particles in the Universe exists.
In fact, the luminous matter can only account for a density $\simeq 0.005$ and the baryonic Dark Matter 
for $\simeq 0.04$. On the other hand, the contribution of Dark Matter particles relativistic at the decoupling
time is also restricted to be $\lsim 0.01$ by considerations on large scale structure formations \cite{CrEl}.
Thus, most of the Dark Matter particles in the Universe, relics from the Big Bang,
were non relativistic at decoupling time; they are 
named Cold Dark Matter particles (CDM). The
CDM candidates have to be neutral, stable or quasi-stable (e.g. with a time decay of order of the age of the Universe)
and have to weakly interact with ordinary matter. These features are respected by 
the axions (also investigated by DAMA/NaI \cite{Ax})
and by a class of candidates named WIMPs (Weakly Interacting Massive Particles).
In particular, in the Standard Model of particle Physics no particle can be a suitable 
candidate as CDM; thus, a new window beyond the Standard Model can be investigated.

Finally, several observations have suggested that 
our Galaxy should also be embedded in a dark halo with mass at least 10 times
larger than that of the luminous matter. The verification of the presence of 
a Dark Matter particle component in the galactic halo has been the main goal of the 
DAMA/NaI experiment.

\section{The DAMA/NaI experiment}
 
\begin{figure}[!ht]
\centering
\includegraphics[height=6.cm]{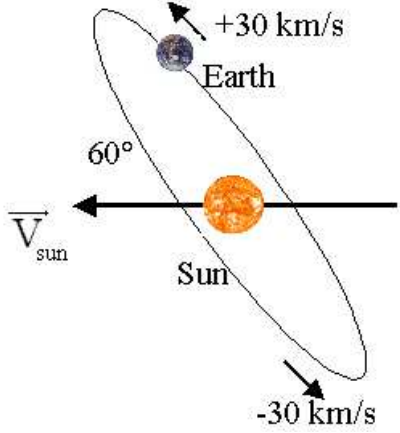}
\caption{Schematic view of the Earth motion around the Sun.}
\label{fg:annmod1}
\end{figure}

The DAMA/NaI experiment was proposed in 1990 \cite{Prop}, designed and realized having the main aim to 
investigate in a model independent way the presence of a Dark Matter particle component in the galactic halo
\cite{Mod1,Mod2,Ext,Mod3,Sist,Sisd,Inel,Hep,RNC}.
For this purpose, we planned to exploit the effect of the Earth revolution around the Sun on the Dark Matter particles
interactions on the target-nuclei of suitable underground detectors. In fact, as a consequence of its
annual revolution, the Earth should be 
crossed by a larger flux of Dark Matter particles in June (when its rotational velocity is summed 
to the one of the solar system with respect to the Galaxy)
and by a smaller one in December (when the two velocities are subtracted) (see
Fig. \ref{fg:annmod1}).
This offers an efficient model independent signature, able to test a large interval of 
cross sections and of halo densities; it is named {\it annual modulation 
signature} and was originally suggested in the middle of '80 by \cite{Freese}. 

In fact, the expected differential rate
as a function of the recoil energy, $dR/dE_R$ (see ref. \cite{RNC} for detailed 
discussion), depends on the WIMP velocity distribution 
and on the Earth's velocity in the galactic frame, $\vec{v}_e(t)$.
Projecting $\vec{v}_e(t)$ on the galactic plane, one can write:
$v_e(t) = v_{\odot} + v_{\oplus} cos\gamma cos\omega(t-t_0)$.
Here $v_{\odot}$ is the Sun's velocity with the respect 
to the galactic halo ($v_{\odot} \simeq v_0 + 12$ km/s and $v_0$ is 
the local velocity whose value is in the range 170-270 km/s \cite{Ext,loc}); 
$v_{\oplus}$ = 30 km/s is the Earth's orbital
velocity around the Sun on a plane with inclination
$\gamma$ = 60$^o$ with the respect to the galactic plane. Furthermore, 
$\omega$= 2$\pi$/T with T=1 year and roughly t$_0$ $\simeq$ 2$^{nd}$ June 
(when the Earth's speed is at maximum). The Earth's velocity can be 
conveniently expressed in unit of
 $v_0$: $\eta(t) = v_e(t)/v_0 = \eta_0 + 
 \Delta\eta cos\omega(t-t_0)$, 
where -- depending on the assumed value of the 
local velocity -- $\eta_0$=1.04-1.07 is the yearly average of $\eta$ and 
$\Delta\eta$ = 0.05-0.09. Since $\Delta\eta\ll\eta_0$, the expected counting 
rate can be expressed by the first order Taylor approximation:
\begin{equation}
\frac{dR}{dE_R}[\eta(t)] = \frac{dR}{dE_R}[\eta_0] +
\frac{\partial}{\partial \eta} \left( \frac{dR}{dE_R} \right)_{\eta =
\eta_0} \Delta \eta \cos\omega(t - t_0) .
\end{equation}
Averaging this expression in a $k$-th energy interval one obtains:
\begin{equation}
	S_k\lbrack\eta(t)\rbrack = S_k\lbrack\eta_0\rbrack
  + \lbrack\frac{\partial  S_k}{\partial \eta}\rbrack_{\eta_0}
\Delta\eta cos\omega(t-t_0) =S_{0,k} + S_{m,k}cos\omega(t-t_0),
\label{eq:sm}
\end{equation}
with the contribution from the highest order terms less than 0.1$\%$.
Thus, the annual modulation signature is very distinctive
since a WIMP-induced seasonal effect must simultaneously satisfy
all the following requirements: (i) the rate must contain a component
modulated according to a cosine function; (ii) with one year period;
(iii) a phase that peaks roughly around $\simeq$ 2$^{nd}$ June;
(iv) this modulation must only be found
in a well-defined low energy range, where WIMP induced recoils
can be present; (v) it must apply to those events in
which just one detector of many actually "fires", since
the WIMP multi-scattering probability is negligible; (vi) the modulation
amplitude in the region of maximal sensitivity must be $\lsim$7$\%$ for usually adopted halo distributions, but it can
be larger in case of some possible scenarios such as e.g. those in refs. \cite{Wei01,Fre04}.
Only systematic effects able to fulfil these 6 requirements and to account for the whole observed modulation amplitude 
could mimic this signature;
thus, no other effect investigated so far in the field of rare processes offers 
a so stringent and unambiguous signature.
With the 
present technology, the annual modulation signature remains the 
main signature of a WIMP signal.

Of course, the amount of the measured effect depends e.g.: i) on the sensitivity of the 
experiment to the coupling and to the particle physics features of the WIMP; ii) on the 
nuclear features of the used tar\-get-nuc\-leus; iii) on the features of the dark halo
and, in particular,  on the spatial and velocity distributions
of the Dark Matter particles.
The quality of the running conditions also plays a crucial role to detect such a rare process.

Considering its main goal, DAMA/NaI has been designed by employing and further developing all the 
necessary low-background techniques and procedures.
A detailed description of the set-up,
of its radiopurity, of its performance, of the used
hardware procedures, of the determination of the experimental quantities 
and of the data reduction has been given in refs. \cite{Nim98,Mod3,Sist,RNC}.
Here only few arguments are addressed.

The DAMA/NaI experiment was located deep underground in the 
Gran Sasso National Laboratory of I.N.F.N. \footnote{We take this occasion to remind that 
DAMA/NaI has been part of the DAMA project, which is also
composed by several other low background set-ups, such as:
i) DAMA/LXe ($\simeq$ 6.5 kg pure liquid Xenon scintillator) \cite{DAMALXe,Xe98};
ii) DAMA/R\&D, set-up devoted to tests on prototypes and small scale experiments
\cite{DAMARD};
iii) the new second generation large mass NaI(Tl) radiopure set-up
DAMA/LIBRA (see later);
iv) DAMA/Ge detector for sample measurements.}, whose main features have been reported 
in \cite{Mac97,Neu89,Cri,Arpe}.

In particular, the NaI(Tl) scintillator was chosen as the best target material to investigate the process since
it offers e.g.: 
\begin{itemize}
\item well known technology
\item reachable high radiopurity by material selections and protocols, by chemical/physical purifications, etc.
\item large mass 
\item high duty cycle 
\item feasible well controlled operational conditions and monitoring
\item routine calibrations feasible down to keV range in the same conditions as the production runs
\item high light response, that is keV threshold reachable
\item absence of the necessity of re-purification or cooling down/warming up procedures (implying high 
      reproducibility, high stability, etc.)
\item absence of microphonic noise and an  effective noise rejection at threshold (time decay of NaI(Tl) pulses is 
      hundreds ns, while that of noise pulses is tens ns)
\item sensitivity to  spin-independent (SI), spin-dependent (SD) and mixed (SI\&SD) couplings as well as to 
      several other existing scenarios
\item sensitivity to both high (by Iodine target) and low (by Na target) mass candidates
\item possibility to effectively investigate the annual modulation signature in all the needed aspects
\item pulse shape discrimination feasible at reasonable level
\item possibility to achieve significant results on several other rare processes
\item no safety problems
\item necessity of a relatively small underground space
\item the lowest cost with the respect to every other considered technique
\end{itemize}

However, neither commercial low background NaI(Tl) detectors 
nor NaI(Tl) detectors grown with old technology (even after "revision") can reach the needed sensitivity; 
thus, a devoted special R\&D was realized with the Crismatec company. 
In this framework, all the needed materials were selected and through
devoted and severe protocols the final DAMA/NaI highly radiopure detectors 
were realized \footnote{We remark that
the low background technique requires very long and accurate work for 
the selection of low radioactive materials by sample
measurements with HP-Ge detectors (placed deep
underground in suitable hard shields) and/or by mass spectrometer analyses.
Thus, these measurements are often 
difficult experiments themselves, depending on the required level of radiopurity.}.

The performances of the nine 9.7 kg highly radiopure DAMA/NaI detectors, 
including their measured radiopurity, are discussed
in details in \cite{Nim98,Sist,RNC}.
The bare NaI(Tl) crystals are 
encapsulated in radiopure Cu housings; moreover,
10 cm long Tetrasil-B light guides act as optical windows on the two 
end faces of each crystals and are 
coupled to specially developed low background photomultipliers (PMT).
The measured light response is 5.5 -- 7.5 photoelectrons/keV
depending on the detector. 
The two PMTs of a detector
work in coincidence with hardware thresholds at the single 
photoelectron level in order to assure high efficiency for the coincidence at few keV level.
The energy threshold of the experiment, 2 keV, is instead determined by means of X-rays sources and 
of keV range Compton electrons
on the basis also of the features of the noise rejection procedures and of the efficiencies when lowering the number of
available photoelectrons \cite{Nim98}.

The detectors are enclosed in a sealed copper box, continuously maintained in HP Nitrogen 
atmosphere in slightly overpressure with respect to the external environment.

A suitable low background
hard shield against electromagnetic and neutron background was realized 
using very high radiopure Cu and Pb bricks \cite{Nim98}, Cd foils and 10/40 cm 
polyethylene/paraffin; the hard shield is also sealed in a plexiglas box 
and maintained in the high purity (HP) Nitrogen atmosphere.
Moreover, about 1 m concrete (made from the Gran Sasso
rock material) almost fully surrounds the hard shield outside the barrack and at its bottom,
acting as a further neutron moderator.

A three-level sealing system from environmental Radon is effective. In fact, the inner part of the
barrack, where the set-up is allocated, has the floor (above the concrete) 
and all the walls sealed by Supronyl (permeability: 2 $\cdot$ 10$^{-11}$ cm$^2$/s \cite{Woi}) 
plastic and the entrance door is air-tight. A low level oxygen alarm 
informs the operator before entering the inner part of the barrack since
the HP Nitrogen which fills both the inner Cu box 
and the external plexiglas box 
is released in this closed environment. The Radon level inside the (sealed) barrack is continuously monitored and recorded 
with the production data \cite{Mod1,Mod2,Nim98,Mod3,Sist,RNC}.

On the top of the shield a glove-box (also maintained in the HP Nitrogen
atmosphere) is directly connected to the 
inner Cu box, housing the detectors, through Cu pipes. The pipes 
are filled with low 
radioactivity Cu bars (covered by 10 cm of low radioactive Cu and 15 cm of low
radioactive Pb) which can be removed to allow the insertion
of radioactive sources for calibrating the detectors in the same 
running condition, without any contact with external environment \cite{Nim98}.

The whole installation is air-conditioned and the operating temperature as well as many other parameters
are continuously monitored and acquired with the production data.
Moreover, self-controlled
computer processes automatically monitor 
several parameters and manage alarms \cite{Nim98,Sist,RNC}.

The electronic chain and the data acquisition system operative up to summer 2000 have been described in ref. 
\cite{Nim98}, while the new electronics and DAQ installed in summer 2000 
have been described in ref. \cite{RNC}. 

As regards other aspects, we recall that
the linearity and the energy resolution of the detectors have been 
investigated using several radioactive sources \cite{Nim98,Sist} such as,
for the low energy region, $^{55}$Fe (5.9 keV X-rays), 
$^{109}$Cd (22 keV X-rays and 88 keV $\gamma$ line) and 
$^{241}$Am (59.5 keV $\gamma$ line) sources.
In particular, in the production runs, the knowledge of 
the energy scale is assured by periodical calibrations with 
$^{241}$Am source and by monitoring (in the production data themselves
summed every $\simeq$ 7 days) the position and energy resolution 
of the 46.5 keV $\gamma$ line of the $^{210}$Pb 
\cite{Nim98,Psd96,Mod1,Mod2,Mod3,Sist,RNC}. The
latter peak was present -- at level of few counts per day per kg (cpd/kg) -- in the 
measured energy distributions mainly
because of a contamination (by environmental Radon) of the external surface of the crystals'
Cu housings, occurred 
during the first period of the underground storage of the detectors.

The only procedure applied to the data regards the noise rejection,
which is particularly efficient since the difference in time decay of the NaI(Tl)
scintillation pulses and of the noise pulses is of order of hundreds ns;
the procedure have been described e.g. in refs. 
\cite{Nim98,Mod3,Sist}\footnote{This procedure assures also the rejection 
of any possible contribution either from afterglows 
(when not already excluded by the dedicated 500 $\mu$s veto time; 
see above) induced by high energy events 
or from any possible $\check{C}$erenkov pulse in the
light guide or in the PMTs; in fact, they also have fast time decay (of 
order of tens ns).}. 

All the periodical long calibration procedures \cite{Nim98,Psd96}
and the time specifically allocated for maintenance and/or for improvements 
are the main components affecting the duty cycle of the experiment.
Moreover, in the first period data have been taken only in the two extreme 
conditions for the annual modulation signature.

The energy threshold, the PMT gain and the electronic line stability are
continuously verified and monitored during the data taking by the 
routine calibrations, by the position and energy resolution of the $^{210}$Pb line (see
above) and by the study of the hardware rate behaviours with time.

Finally, we remind that the experiment took data up to MeV energy region 
despite the optimization was done for the keV energy range.

The measured low energy distributions of interest for the WIMP investigation have been given 
in refs. \cite{Diu99,Mod3,Sist,Ela99}, where  
the corrections for 
efficiencies and acquisition dead time have already been applied.
We note that usually in DAMA/NaI the low energy distributions refer to 
those events where only one detector of many actually fires
(that is, each detector has all the 
others in the same installation
as veto; this assures a further background 
reduction, which is of course impossible when a single detector is used), {\it single-hit} events.

Moreover, the data taking of each annual cycle has been started before the
expected minimum
of the signal rate (which is roughly around $\simeq$ 2$^{nd}$
December) and concluded after the expected maximum (which is
roughly around $\simeq$ 2$^{nd}$ June).

The DAMA/NaI set-up has exploited the WIMP annual modulation signature over seven annual cycles
\cite{Mod1,Mod2,Ext,Mod3,Sist,Sisd,Inel,Hep,RNC} and the following part of this paper will summarize the final model
independent result, some of the corollary quests for the candidate particle, implications and pespectives. However, it is worth 
to remind that - thanks to its radiopurity and features - DAMA/NaI has also investigated other approaches for WIMPs 
in ref. \cite{Psd96,Diu99} and several other rare processes such as:
possible processes violating the Pauli exclusion principle \cite{naipep}, CNC processes 
in $^{23}$Na and $^{127}$I \cite{naicnc}, electron stability \cite{Ela99}, 
searches for neutral SIMPs \cite{Besimp}, for neutral nuclearites \cite{Besimp}, 
for Q-balls \cite{qballs} and for solar axions \cite{Ax}; 
moreover, DAMA/NaI allowed also the study of nuclear transition to a possible
superdense nuclear state \cite{dama_super}
and possible two-body cluster decay of $^{127}$I \cite{dama_clus}.

\section{The 6.3 $\sigma$ C.L. model-independent evidence for a WIMP component in the galactic halo}

As reported in ref. \cite{RNC}, a model independent investigation of the annual modulation
signature has been realized by exploiting the time behaviour of the
residual rates of the {\it single-hit} events in the lowest energy regions 
over the seven annual cycles (total exposure: 107731 kg $\cdot$ day) \cite{Mod1,Mod2,Ext,Mod3,Sist,Sisd,Inel,Hep,RNC}. 
These experimental residual rates of the {\it single-hit} events
are given by: $<r_{ijk} - flat_{jk}>_{jk}$, where $r_{ijk}$ is the rate
in the considered $i$-th time interval for the $j$-th detector in
the $k$-th considered energy bin and
$flat_{jk}$ is the rate of the $j$-th detector in the $k$-th energy bin averaged
over the cycles. The average is made on all the detectors ($j$ index) and on
all the energy bins in the considered energy interval.

\begin{figure}[!hb]
\begin{center}
\vspace{-1.0cm}
\epsfig{file=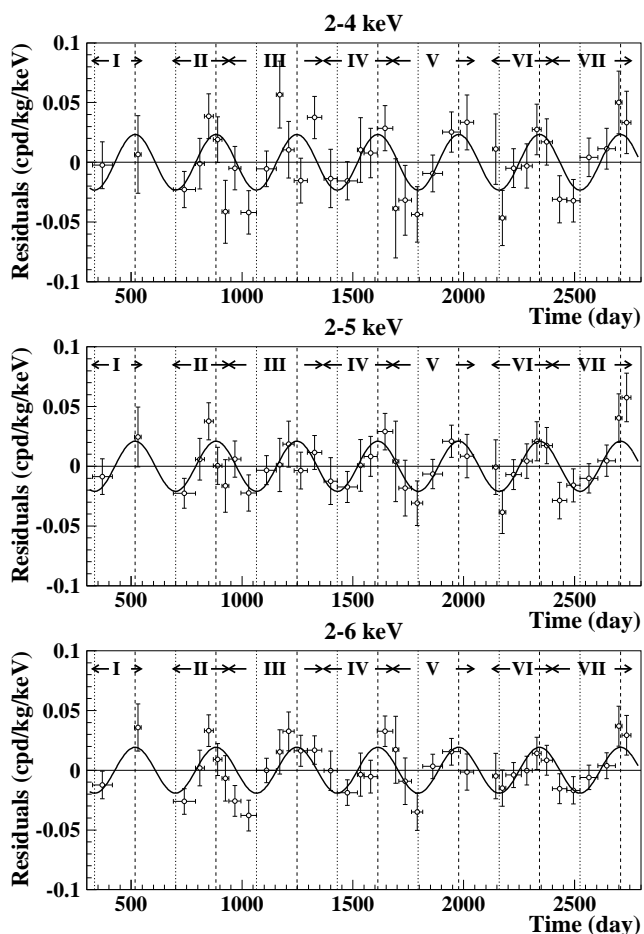,height=13cm}
\end{center}
\vspace{-1.0cm}
\caption{Experimental residual rate for {\it single-hit} events, in the (2--4), (2--5) 
and (2--6) keV energy intervals as a function of the time elapsed since
January 1-st of the first year of data taking. The
experimental points present the errors as vertical
bars and the associated time bin width as horizontal bars. The 
superimposed curves represent the cosinusoidal functions 
behaviours expected for a WIMP signal 
with a period equal to 1 year and phase at $2^{nd}$ June; the modulation
amplitudes have been obtained by best fit. See text.
The total exposure is 107731 kg $\cdot$ day.}
\label{fig2}
\vspace{-0.2cm}
\end{figure}

This model independent approach 
offers an immediate evidence of the presence of 
an annual modulation of the rate of the {\it single-hit} events
in the lowest energy region over the seven annual cycles
as shown in Fig. \ref{fig2}, where the energy and time behaviours of the 
{\it single-hit} residual rates are depicted. 

In particular, the data favour the presence of a modulated cosine-like behaviour 
at 6.3 $\sigma$ C.L.; in fact, 
their fit for the (2--6) keV 
larger statistics energy interval offers a modulation amplitude equal to
$(0.0200 \pm 0.0032)$ cpd/kg/keV, a phase:
$t_0 = (140 \pm 22)$ days and a period:
$T = (1.00 \pm 0.01)$ year, all parameters kept free in the fit.
The period and phase agree with those expected in the case of a
WIMP induced effect ($T$ = 1 year and $t_0$ roughly at $\simeq$ 152.5-th day of the year). 
The $\chi^2$ test on the (2--6) keV residual rate in Fig. \ref{fig2}
disfavours the hypothesis of unmodulated behaviour giving a probability of $7 \cdot
10^{-4}$ ($\chi^2/d.o.f.$ = 71/37).
Moreover, if the experimental residuals of Fig. \ref{fig2} are fitted fixing 
the period at 1 year and the phase at $2^{nd}$ June, the following modulation
amplitudes are obtained: $(0.0233 \pm 0.0047)$ cpd/kg/keV, $(0.0210 \pm 0.0038)$
cpd/kg/keV and $(0.0192 \pm 0.0031)$ cpd/kg/keV 
for the (2--4), (2--5) and (2--6) keV energy intervals, respectively.

\begin{figure}[!ht]
\centering
\epsfig{file=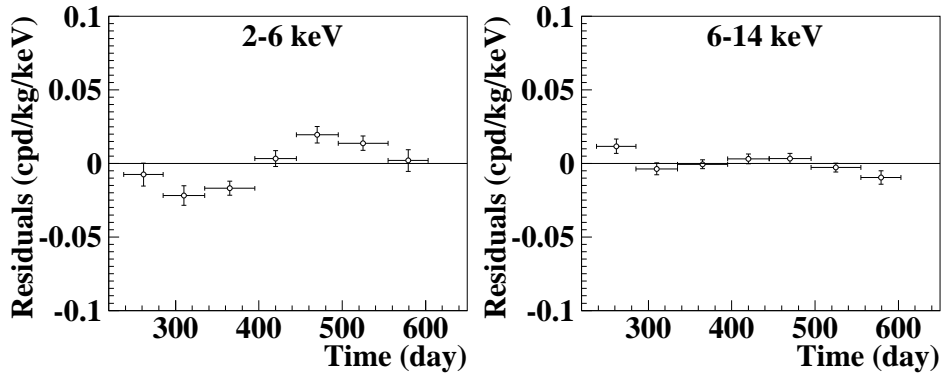,height=6.cm}
\vspace{-1.cm}
\caption{Experimental {\it single-hit} residual rate 
in a single annual cycle from the total exposure of
107731 kg $\cdot$ day for the (2--6) keV and (6--14) keV energy intervals. The
experimental points present the errors as vertical
bars and the associated time bin width as horizontal bars.
The initial time is taken at August 7$^{th}$.
Fitting the data with a cosinusoidal function with period of 1 year and phase at 152.5 days,
the following amplitudes are obtained: $(0.0195 \pm 0.0031)$ cpd/kg/keV and $-(0.0009 \pm 0.0019)$
cpd/kg/keV, respectively. Thus, a clear modulation is present in the lowest energy region, while it
is absent just above.}
\label{fig3}  
\end{figure}

\begin{figure}[!ht]
\begin{center}
\epsfig{file=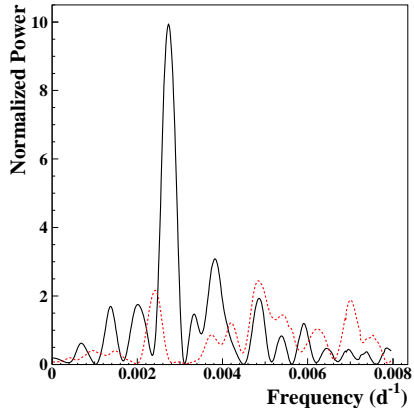,height=5.5cm}
\end{center}
\vspace{-0.7cm}
\caption{Power spectrum of the measured {\it single-hit} residuals
for the (2--6) keV (continuous line) and (6--14) keV (dotted line)
energy intervals calculated according to ref. 
\cite{Lomb}, including also the treatment of the experimental errors and of the 
time binning. As it can be seen, the principal mode present in the (2--6) keV 
energy interval corresponds to a frequency of $2.737 \cdot 10^{-3}$ d$^{-1}$,
that is to a period of $\simeq$ 1 year. A similar peak is not present in the 
(6--14) keV energy interval.}
\label{fig4}
\normalsize
\end{figure}

In Fig. \ref{fig3} the {\it single-hit} residual rate in a single annual cycle 
from the total exposure of 107731 kg
$\cdot$ day is presented for two different energy intervals; as it
can be seen the modulation is clearly present in the (2--6) keV energy region, while 
it is absent just above.
The same conclusion is obtained by investigating the data by means of the Fourier analysis (performed according to ref.
\cite{Lomb} including also the treatment of the experimental errors and of the time binning); in fact,
the results shown in Fig.~\ref{fig4} show 
a clear peak for a period of 1 year in the (2--6) keV energy interval, while 
it is absent in the energy interval just above.

A quantitative investigation of the whole energy spectrum up to MeV energy region has
not shown modulation in any other energy interval (see e.g. \cite{Mod3,Sist,RNC}
and few arguments given later).

\begin{figure}[!ht]
\begin{center}
\mbox{\epsfig{file=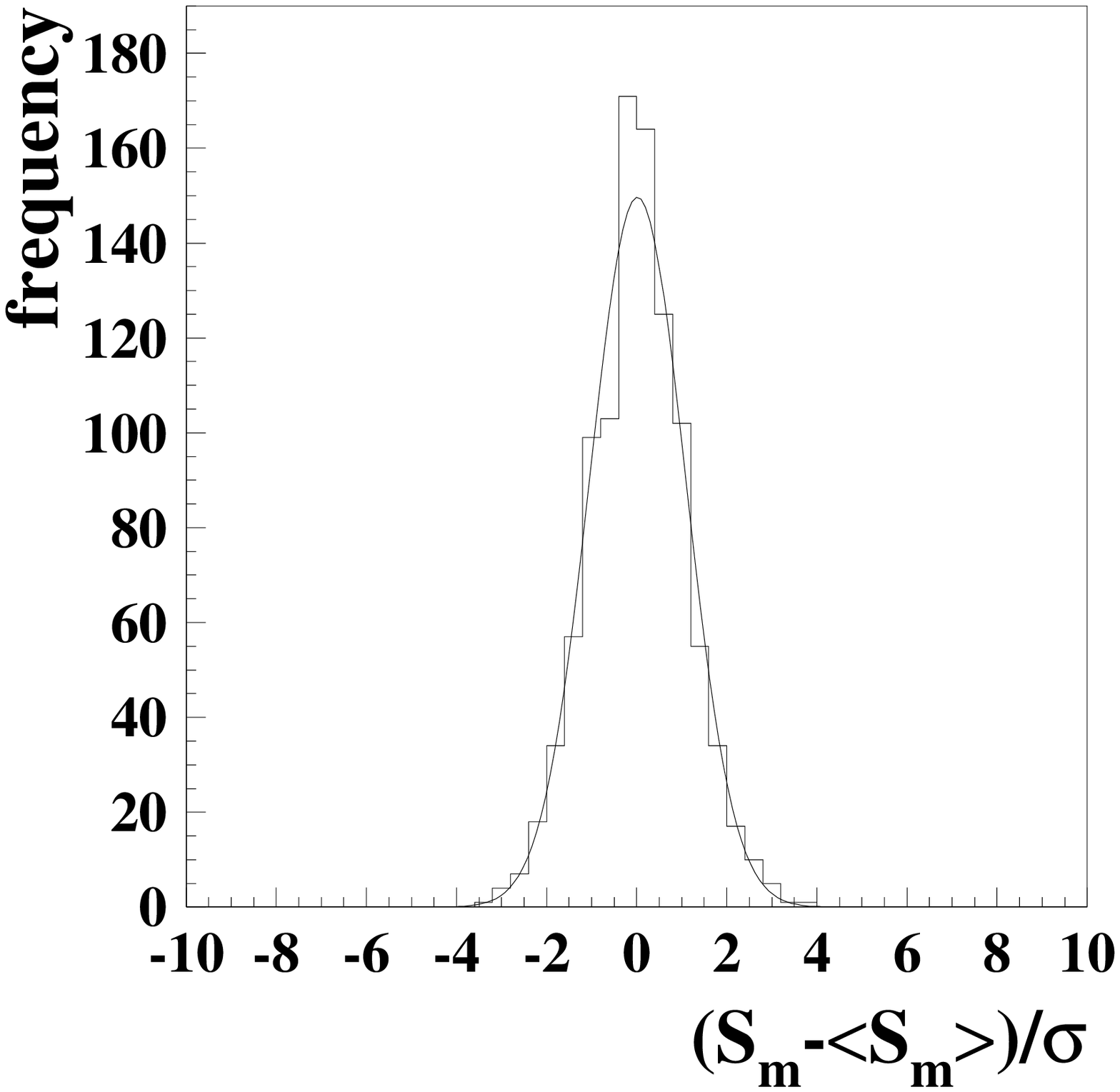,height=6.0cm}}
\mbox{\epsfig{file=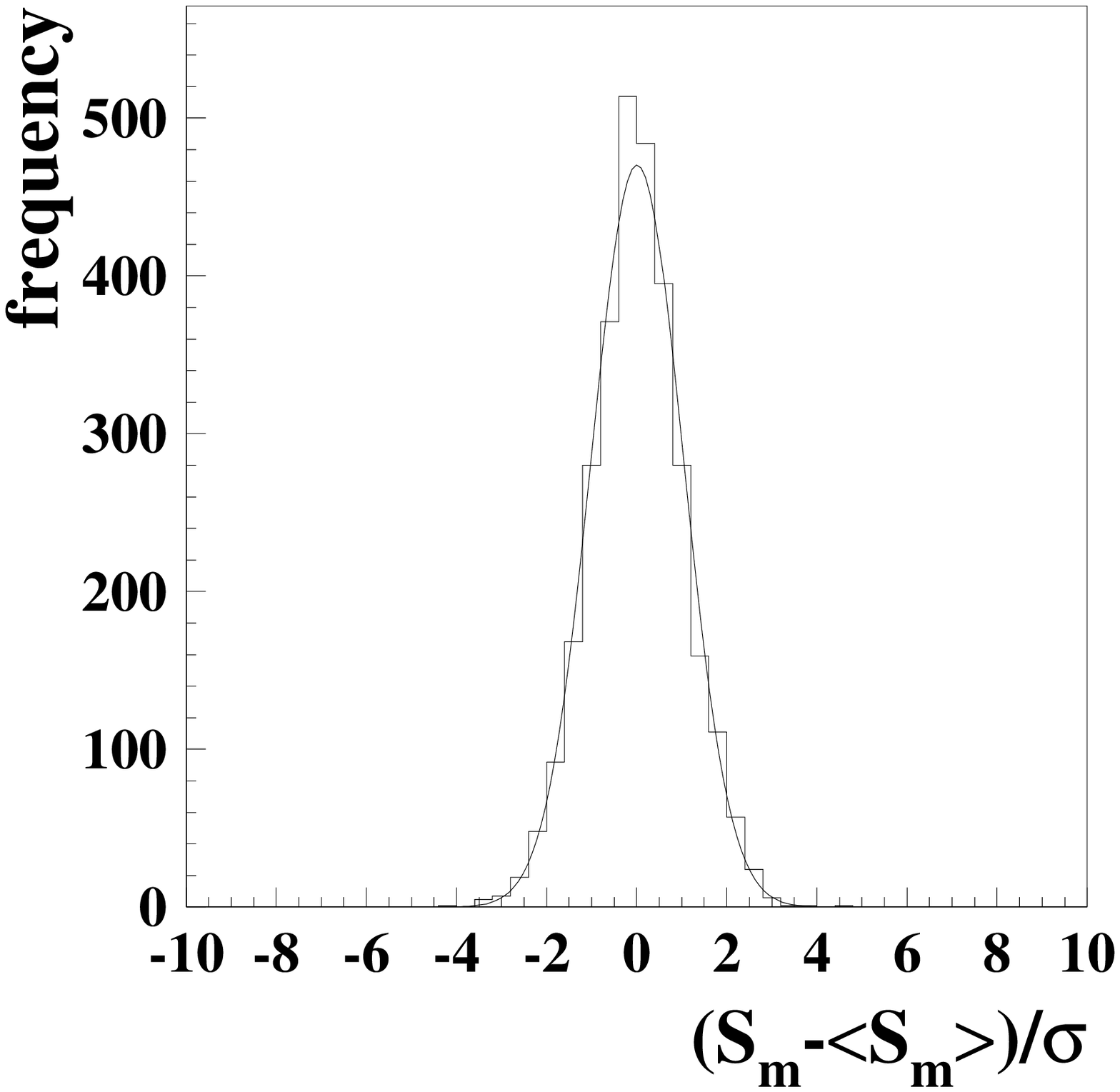,height=6.0cm}}
\end{center}
\vspace{-1cm}
\caption{Distributions of the variable $\frac {S_m-<S_m>}{\sigma}$ (where $\sigma$ is the error
associated to the $S_m$) evaluated for each detector,   
for each annual cycle and each considered energy bin: i) in the region of interest for   
the observed modulation, (2--6)
keV (left panel);
ii) including also the energy region just above, (2--14) keV (right panel). See text.}
\label{gaus1}
\end{figure}

Finally, in order to show if the modulation amplitudes are statistically 
well distributed in all the crystals, in all the annual cycles and  
in the energy bins, the distributions of the variable $\frac {S_m-<S_m>}{\sigma}$
are reported in Fig. \ref{gaus1}. The $S_m$ are the
experimental modulation amplitudes for each detector,
for each annual cycle and for each considered energy bin (taken
as an example equal to 0.25 keV), $\sigma$ are their
errors and the $<S_m>$ represent the mean values of the modulation amplitudes over
the detectors and the annual cycles for each energy bin.
The left panel of Fig. \ref{gaus1} shows the distribution referred to the
region of interest for the observed modulation, (2--6)
keV, while the right panel includes also the energy region just above, (2--14) keV. 
Since this variable is distributed as a 
gaussian with an unitary standard deviation, the
modulation amplitudes are statistically well distributed in all the crystals,
in all the data taking periods and in all
considered energy bins.

In conclusion, the data satisfy all the peculiar requirements, given above,
for the WIMP model independent annual modulation signature.
As mentioned, to mimic this
signature, systematic effects or side reactions should not only be able to 
fully account for the measured modulation amplitude,
but also to satisfy the
peculiar requirements as for a WIMP induced effect.
A careful investigation of all the known possible sources of systematics and side
reactions has been regularly carried out by DAMA/NaI and published at time of each data 
release \cite{Mod1,Mod2,Mod3,Sist,RNC}.

No effect able to mimic the signature has been found;
thus, upper limits (90\% C.L.)  
on the possible contributions to the modulated amplitude
have been calculated as summarized in Table \ref{tb:sist};
for a detailed quantitative discussion see ref. \cite{Sist,RNC}.
\begin{table}[ht]
\caption{Summary of the results obtained by investigating the possible sources of 
systematics or of side reactions \cite{RNC}.
No systematics or side reaction has been found able to give a modulation amplitude 
different from zero;
thus cautious upper limits (90\% C.L.)
have been calculated and are shown here
in terms of the measured model independent modulation amplitude, $S_m^{obs}$.
As it can be seen, none of them nor
their cumulative effect could account for the 
measured modulation; moreover, as discussed in details in ref. \cite{Sist,RNC},
they cannot mimic the signature.}
\begin{center}
\begin{tabular}{|c|c|}
\hline \hline
 Source    &  Cautious upper limit \\
           & (90\%C.L.)            \\
\hline\hline
Radon      & $<0.2\% S_m^{obs}$    \\
\hline
Temperature & $<0.5\% S_m^{obs}$   \\
\hline
Noise      & $<1\% S_m^{obs}$      \\
\hline
Energy scale & $<1\% S_m^{obs}$    \\
\hline
Efficiencies & $<1\% S_m^{obs}$    \\
\hline
Background   & $<0.5 \% S_m^{obs}$ \\
\hline
Side reactions & $<0.3\% S_m^{obs}$  \\
\hline
\multicolumn{2}{|c|} {In addition: no effect can mimic the signature} \\
\hline \hline
\end{tabular}
\end{center}
\label{tb:sist}
\end{table}
In particular, they cannot account for the
measured modulation amplitude because quantitatively not relevant and
unable to satisfy all the peculiar requirements of the signature.

For the sake of completeness, we remind that possible diurnal 
effects -- correlated both with the sidereal and with the solar time -- 
have already been excluded by the analysis reported in  \cite{Diu99}.

In particular, as mentioned above, no modulation has been observed
in the background; in fact, the whole energy spectrum up to MeV energy region
has been analyzed and the presence of a background
modulation in the whole energy spectrum has been excluded at a level much
lower than the effect found in the lowest energy region \cite{Sist,RNC}.
This result already accounts 
also for the background component due to the neutrons and the Radon;
nevertheless, further additional 
independent and cautious analyses to estimate their possible contribution
have been given in ref. \cite{Sist,RNC}.
In fact, it has been demonstrated in \cite{Sist,RNC} that
a modulation of neutron flux -- possibly observed by the ICARUS coll.
as reported in the ICARUS internal report TM03-01 --
cannot quantitatively
contribute to the DAMA/NaI observed modulation amplitude,
even if the neutron flux would
be assumed to be 100 times larger than measured at LNGS by several authors
with different techniques over more than 15 years.
Moreover, in no case the neutrons can mimic the signature since
some of the peculiar requirements of the signature would fail.
Similarly, any possible effect of the muon flux modulation reported 
by the MACRO experiment \cite{Mac97} is excluded both by quantitative investigation \cite{Sist,RNC}
and by inability to fulfil all the peculiarities of the signature.

As regards the possibility of a contribution from the Radon, we remind that
the DAMA/NaI has three levels of insulation from the environmental air
(see above). Moreover, the Radonmeter which continuosly recorded the Radon level inside the barrack 
(that is external to the hard shield and to the detectors)
typically measured values at level of its sensitivity and no modulation has been observed
\cite{Mod1,Mod2,Mod3,Sist,RNC}.
To be on the safest side, even the possible presence of Radon trace in
the HP Nitrogen atmosphere inside the Cu box, housing the detectors, has been investigated 
by searching 
for the double coincidences of
the gamma-rays (609 and 1120 keV) from $^{214}$Bi
Radon daughter, obtaining an upper 
limit of:
$< 4.5 \cdot 10^{-2}$ Bq/m$^3$  (90\% C.L.). 
It gives rise to the upper limit reported in Table \ref{tb:sist}
when assuming an hypothetical 10\%, modulation of possible Radon in 
the HP Nitrogen atmosphere of the Cu box.
Anyhow, it is worth to remark that
in every case even a sizeable quantity 
of Radon nearby a detector cannot mimic the WIMP 
annual modulation signature since 
some of the peculiarities of the signature would fail.

For more detailed discussion see ref. \cite{Sist,RNC}.

To perform a further relevant investigation, in 1999 we proposed 
to renew the electronic chain of DAMA/NaI removing the multiplexer system and equipping each detector with its
own transient digitizer. This occurred in summer 2000, thus in the last two annual cycles
the {\it multiple-hits} events and the {\it single-hit} events have been acquired and analyzed using the same 
identical hardware and the same identical software procedures.
\begin{figure}[!ht]
\begin{center}
\vspace{-1cm}
\mbox{\epsfig{file=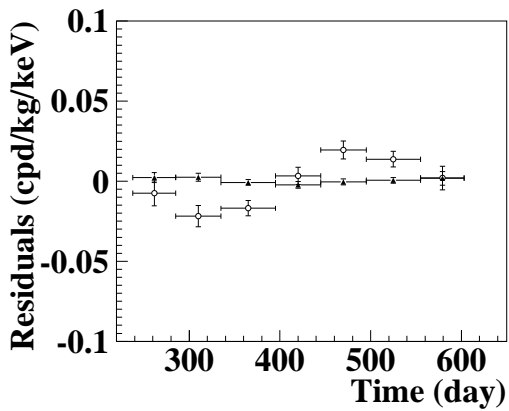,height=6.0cm}}
\end{center}
\vspace{-0.5cm}
\caption[]{Experimental residual rates over seven annual cycles
for {\it single-hit} events (open circles) -- class of events 
to which WIMP events belong --
and over the last two annual cycles for {\it multiple-hits} 
events (filled triangles)
-- class of events to which WIMP events do not belong -- in the
(2--6) keV cumulative energy interval. They have been obtained 
by considering for each class of events the data as collected 
in a single annual cycle and using in both cases the same 
identical hardware and the same identical software
procedures. The initial time is taken on August 7$^{th}$.
See text.}
\label{fig_mul}
\end{figure}
The {\it multiple-hits} events class -- on the contrary of the {\it single-hit} one
-- does not include events induced by
WIMPs since the probability that a WIMP scatters 
off more than one detector is
negligible.  
The obtained results are depicted in Fig. \ref{fig_mul}.
The fitted modulation amplitudes are: $A=(0.0195\pm0.0031)$ cpd/kg/keV 
and $A=-(3.9\pm7.9) \cdot 10^{-4}$ cpd/kg/keV for {\it single-hit} 
and {\it multiple-hits} residual rates, respectively.
Thus, evidence of annual modulation is present in 
the {\it single-hit} residuals (events class to which the WIMP-induced recoils
belong), while it is absent in the 
{\it multiple-hits} residual rate (event class to which only background events belong).
Since the same identical hardware and the same identical software procedures have been 
used to analyse the two classes of events,
the obtained result offers an additional strong support for the presence of 
a Dark Matter particle component in the galactic halo further excluding any side effect 
either from hardware or from software procedures or from background.

In conclusion, the presence of a Dark Matter particle component 
in the galactic halo is supported by DAMA/NaI at 6.3 $\sigma$ C.L.
and the modulation amplitude measured over the 7 annual cycles in NaI(Tl)
at the location of the Gran Sasso Laboratory 
for the (2 -- 6) keV energy region is 
$(0.0200 \pm 0.0032)$ cpd/kg/keV.
This is the experimental result of DAMA/NaI. It is model
independent; no other experiment whose result can be
directly compared with this one 
is available so far in the field of Dark Matter
investigation.

\section{Corollary results: quests for a candidate particle in some of the possible model
frameworks}

On the basis of the obtained model independent result, corollary 
investigations can also be
pursued on the nature and coupling of the WIMP candidate.
This latter investigation is instead model
dependent and -- considering the large uncertainties which exist on the 
astrophysical, nuclear and particle physics
assumptions and on the parameters needed in the calculations -- has no general
meaning (as it is also the case of exclusion plots and 
of the WIMP parameters evaluated in the indirect
detection experiments). Thus, it should be
handled in the most general way as we have preliminarily
pointed out with time passing in the past
\cite{Mod1,Mod2,Ext,Mod3,Sist,Sisd,Inel,Hep} and we have discussed
in some specific details in ref. \cite{RNC}.

It is worth to note that the results presented in the following 
are, of course, not exhaustive of the many possible scenarios
which at present level of knowledge 
cannot be disentangled. Some of the open questions are: i) which is 
the right nature for the WIMP particle;
ii) which is its right couplings with ordinary matter (mixed SI\&SD, 
purely SI, purely SD or {\em preferred inelastic});
iii) which are the right form factors and
related parameters for each target nucleus; iv) 
which is the right spin factor for each target nucleus;
v) which are the right scaling laws (let us remind that 
even for the neutralino case in a MSSM framework with purely SI interaction the scenario could be
drastically modified as pointed out in ref. \cite{Kam03}); 
vi) which is the right halo model and
related parameters; vii) which are the right values 
of the experimental parameters within their uncertainties; etc.

As regards, in particular, the Dark Matter particle-nucleus elastic scattering, 
the differential energy distribution of the recoil nuclei
can be calculated \cite{Psd96,Boat,RNC}
by means of the differential cross section of the WIMP-nucleus elastic 
processes given by the sum of the SI and the SD contributions. 
In the SI case, the nuclear parameters can be decoupled from the
particle parameters and the nuclear cross sections are usually scaled to a defined 
point-like SI Dark Matter particle-nucleon cross section, $\sigma_{SI}$.  
In the SD case the notations \cite{Sisd}:
$tg \theta = \frac{a_n}{a_p}$, can be used, 
where $a_{p,n}$ are the effective WIMP-nucleon coupling strengths for SD interactions. 
The mixing angle $\theta$ is defined in the $\left[ 0, \pi \right)$ interval;
in particular, $\theta$ values in the second sector 
account for $a_p$
and $a_n$ with different signs. Also in the SD case all the nuclear cross sections 
are usually scaled to a defined             
point-like SD Dark Matter particle-nucleon cross section, $\sigma_{SD}$ \cite{RNC}.
Thus, the energy distribution of the recoil rate can be written as a function of 
$\sigma_{SI}$, $\sigma_{SD}$ and $\theta$. Another important parameter is 
the local density  of the Dark Matter particles, $\rho_W = \xi \rho_0$, 
where $\rho_0$ is the local halo density and $\xi$ ($\xi\leq 1$)
is the fractional amount of local density of Dark Matter particles.

Among the ingredients entering in the model dependent analyses 
there are the nuclear SI and SD form factors, which 
generally depend on the nature of the involved particle.
In ref. \cite{RNC} the existing uncertainties in the usually adopted formulation for the SI case
have been discussed as well as those for the SD one. It has also shown that the SD case is even more uncertain
since the nuclear and particle physics degrees of freedom cannot be decoupled and a dependence 
on the assumed nuclear potential exists. Moreover, 
further significant uncertainties in the evaluation of the SD interaction rate also arise 
from the adopted spin factor for the single target-nucleus. 
In fact, the available calculated values are well different in different models and, in addition,
at  fixed model they depend on $\theta$. Thus,
not only the target nuclei should have spin different from zero (for example, this is not the case of Ar isotopes)
to be sensitive to Dark Matter particles  with a SD component in the coupling, but also
well different sensitivities can be expected among 
odd-nuclei having an unpaired proton (as e.g. $^{23}$Na and $^{127}$I)
and odd-nuclei having an unpaired neutron (as e.g. the odd Xe and Te isotopes and $^{73}$Ge).

Another scenario also considered in the corollary DAMA/NaI model dependent analyses is that of
WIMPs with {\it preferred inelastic} scattering which 
has been suggested by \cite{Wei01}. In this case the Dark Matter particles 
can only inelastically scatter off nuclei going to excited levels with
a $\delta$ mass splitting. A specific model featuring a real component of the sneutrino
(for which the mass splitting naturally arises) has been given in ref. \cite{Wei01}.
It has been shown that for this inelastic scattering 
a kinematical constraint exists which favours
target-detectors media with heavy nuclei (such as $^{127}$I) with the respect to 
those with lighter ones (such as e.g. $^{nat}$Ge).
In fact, this process can only occur if the particle velocity is larger than a threshold value; this 
kinematical constraint becomes increasingly severe as the nucleus mass decreases \cite{Wei01}. 
Moreover, this model scenario implies some interesting peculiar features when 
exploiting the annual modulation signature; in fact -- with the respect to the
case of WIMP elastically scattering -- it would give rise to an enhanced 
modulated component of the signal with respect to the unmodulated one
and to largely different behaviours with energy 
for both the components (showing both higher mean values) \cite{Wei01}.
The differential energy distribution of the recoil nuclei
in the case of inelastic processes is function of $\xi\sigma_p$, $m_W$ and $\delta$,
analogoulsy as above other ingredients, as e.g. the form factor, play also a role \cite{RNC}. 

\vspace{0.3cm}
As mentioned, the expected energy distribution for the scatterings of Dark Matter particles
depends on $\rho_W$ and on the 
velocity distribution of the Dark Matter particles at Earth's position.
The experimental observations regarding the dark halo of our Galaxy
do not allow to get information on them without introducing 
a model for the Galaxy matter density. 
The dark halo model widely used in the calculations carried out in the field of 
direct detection approaches is the naive isothermal sphere 
that corresponds to a spherical infinite system with a flat rotational
curve. Despite its simplicity has favoured its wide use 
in the calculation of expected rate of Dark Matter particle-nucleus  
interaction, it doesn't match with astrophysical observations and presents an unphysical behaviour.
In fact, the density profile has a singularity in the origin and
implies a total infinite mass of the halo unless introducing 
some cut-off at large radii. 
Thus, the use of more realistic halo models is mandatory since the model dependent results
significantly vary. 
An extensive discussion about 
some of the more credited realistic halo models has been reported in ref. \cite{Hep,RNC} and have been considered
in the results given in the following (see Table \ref{tb:halo}).
In particular, the considered halo model classes correspond to: i) spherically symmetric matter density with
isotropic velocity dispersion (A); ii) spherically symmetric matter density with
non-isotropic velocity dispersion (B); iii) axisymmetric models (C);
iv) triaxial models (D);
v) moreover, in the case of axisymmetric models it is possible to include either an halo co-rotation 
or an halo counter-rotation.
\begin{table}[!hbt]
\begin{center}
\vspace{-0.7cm}
\caption{\label{tb:models}
  Summary of the considered consistent halo models \cite{Hep,RNC}.
  The labels in the first column identify the models. 
  In the third column the values of the related considered parameters are reported \cite{Hep,RNC};
  other choices are also possible as well as other halo models.
  The models of the Class C have also been considered including
  possible co--rotation and counter-rotation of the dark halo.
}
\begin{tabular}{|c|l|c|}
\hline\hline
\multicolumn{3}{|l|}{{\bf Class A:  spherical $\bf \rho_{W}$,
isotropic velocity dispersion}} \\
\hline
A0 & {\rm ~Isothermal Sphere}   &     \\
A1 & {\rm ~Evans' logarithmic}  & $R_c=5$ kpc \\
A2 & {\rm ~Evans' power-law}  & $R_c=16$ kpc, $\beta=0.7$ \\
A3 & {\rm ~Evans' power-law}  & $R_c=2$ kpc, $\beta=-0.1$ \\
A4 & {\rm ~Jaffe}               & $\alpha=1$, $\beta=4$,
$\gamma=2$, $a=160$ kpc \\ 
A5 & {\rm ~NFW}                  & $\alpha=1$, $\beta=3$,
$\gamma=1$, $a=20$ kpc \\ 
A6 & {\rm ~Moore et al.}    & $\alpha=1.5$, $\beta=3$,
$\gamma=1.5$, $a=28$ kpc  \\ 
A7 & {\rm ~Kravtsov et al.} & $\alpha=2$, $\beta=3$,
$\gamma=0.4$, $a=10$ kpc   \\
\hline
\multicolumn{3}{|l|}{{\bf Class B: spherical $\bf \rho_{W}$,
non--isotropic velocity dispersion    }} \\
\multicolumn{3}{|l|}{{\bf (Osipkov--Merrit, $\bf \beta_0=0.4$)}} \\
\hline
B1 & {\rm ~Evans' logarithmic} & $R_c=5$ kpc \\
B2 & {\rm ~Evans' power-law} & $R_c=16$ kpc, $\beta=0.7$  \\
B3 & {\rm ~Evans' power-law} & $R_c=2$ kpc, $\beta=-0.1$  \\
B4 & {\rm ~Jaffe}           & $\alpha=1$, $\beta=4$,
$\gamma=2$, $a=160$ kpc  \\
B5 & {\rm ~NFW}             & $\alpha=1$, $\beta=3$,
$\gamma=1$, $a=20$ kpc   \\
B6 & {\rm ~Moore et al.}    & $\alpha=1.5$, $\beta=3$,
$\gamma=1.5$, $a=28$ kpc   \\
B7 & {\rm ~Kravtsov et al.} &  $\alpha=2$, $\beta=3$,
$\gamma=0.4$, $a=10$ kpc    \\
\hline
\multicolumn{3}{|l|}{{\bf Class C:  Axisymmetric $\bf \rho_{W}$}} \\
\hline
C1 & {\rm ~Evans' logarithmic} & $R_c=0$, $q=1/\sqrt{2}$ \\
C2 & {\rm ~Evans' logarithmic} & $R_c=5$ kpc, $q=1/\sqrt{2}$ \\
C3 & {\rm ~Evans' power-law} & $R_c=16$ kpc, $q=0.95$, $\beta=0.9$ \\
C4 & {\rm ~Evans' power-law} & $R_c=2$ kpc, $q=1/\sqrt{2}$, $\beta=-0.1$\\
\hline
\multicolumn{3}{|l|}{{\bf Class D: Triaxial $\bf \rho_{W}$ 
  ($\bf q=0.8$, $\bf p=0.9$)}} \\
\hline
D1 & {\rm ~Earth on maj. axis, rad. anis.}    & $\delta=-1.78$  \\
D2 & {\rm ~Earth on maj. axis, tang. anis. }    &   $\delta=16$ \\
D3 & {\rm ~Earth on interm. axis, rad. anis.}  &  $\delta=-1.78$ \\
D4 & {\rm ~Earth on interm. axis, tang. anis.} & $\delta=16$ \\
\hline\hline
\end{tabular}
\label{tb:halo}
\end{center}
\vspace{-0.5cm}
\end{table}
The parameters of each halo model have been chosen 
taking into account the available observational data.
Thus, considering the allowed range for the local velocity of Dark Matter particles
$v_0=(220 \pm 50)$ km s$^{-1}$ (90\% C.L.),
the allowed range of local density $\rho_0$ has been evaluated \cite{Hep}
taking into account the following physical constraints: i) the amount of flatness of the rotational
curve of our Galaxy, considering conservatively $0.8 \cdot v_0  \lsim
v_{rot}^{100}  \lsim  1.2 \cdot v_0$, where $v_{rot}^{100}$ is the
value of rotational curve at distance of 100 kpc from the galactic
center; ii) the maximal non dark halo components in the Galaxy, considering
conservatively $1 \cdot 10^{10} M_{\odot}  \lsim  M_{vis}  \lsim  6
\cdot 10^{10} M_{\odot}$ \cite{Deh98,Gat96}. 
Although  a large number
of self-consistent galactic halo models, in which the variation of the
velocity distribution function is originated from the change of the 
halo density profile or of the potential, have been considered, still many other possibilities exist.

The proper knowledge of other quantities is also necessary 
such as e.g. the recoil/electron response ratio
for the given nucleus in the given detector and energy range, (named {\em quenching factor}).
Of course, significant differences are often present
in literature for the measured value of this recoil /electron response
ratio even for the same nucleus in the same kind of detector as shown in ref. \cite{RNC}.

\vspace{0.5cm}

In conclusion, just as a corollary of the model independent result
over the seven annual cycles, in the following some of the many possible model dependent quests for a Dark Matter  candidate
are summarized. They have been obtained by
considering the halo models previously mentioned for three of the
possible values 
of the local velocity $v_0$: 170 km/s, 220 km/s and 270 km/s and for the halo 
density values as in the prescriptions of ref. \cite{Hep,RNC}.
The escape velocity has been maintained at the fixed value: 650 km/s.
Of course, it is worth to note that the
present existing uncertainties affecting the knowledge of the escape velocity will significantly
extend allowed regions e.g. in the cases of {\em
preferred inelastic} WIMPs and of light mass 
WIMP candidates; its effect would be instead marginal at large WIMP masses.
All these scenarios have been investigated in some
discrete cases either considering
the mean values of the parameters of the used nuclear form factors and 
of the measured quenching factors (case $A$) or adopting the same procedure as in
refs. 
\cite{Sisd,Inel}. The latter one has been obtained by varying either: i) the mean values of the 
measured $^{23}$Na and
$^{127}$I quenching factors \cite{Psd96} 
up to +2 times the errors; ii) the nuclear radius, $r_n$,
and the nuclear surface thickness parameter, $s$, in the SI Form 
Factor \cite{Helm56} from their central values down to -20\%;
iii)  the $b$ parameter in the considered SD form factor
from the given value \cite{res97} down to -20\% (case $B$). Moreover, we have also considered 
one of the possible more extreme cases where the Iodine nucleus
parameters are fixed at the values of case $B$, while 
for the Sodium nucleus one considers:
i) $^{23}$Na quenching factor at the lowest value measured in literature; 
ii) the nuclear radius, $r_n$,
and the nuclear surface thickness parameter, $s$, in the SI Form
Factor \cite{Helm56} from their central values up to +20\%;
iii)  the $b$ parameter in the considered SD form factor
from the given value \cite{res97} up to +20\%
(case $C$). 

Finally, no restriction on the mass of the Dark Matter particle has been adopted in these analyses;
hence, we have just marked on those figures the lowest bound on the
neutralino mass as derived from the LEP data in the adopted supersymmetric
schemes based on GUT assumptions \cite{Dpp0}. In fact,  
other model assumptions are possible and 
would imply significant variations of some accelerators bounds, allowing neutralino mass down to 6 GeV 
(see e.g. the recent refs. \cite{Bo03,lowm,bbpr}); in addition,
other low mass candidates can be considered as well.  
It is worth to note that the LEP model dependent  mass limit -- when considered -- selects the 
WIMP-Iodine elastic scatterings as dominant because of the adopted
scaling laws and of kinematical arguments, while  
DAMA/NaI is
intrinsically sensitive both to low and high WIMP mass having both a light
(the $^{23}$Na) and a heavy (the $^{127}$I) target-nucleus.

The results presented by DAMA/NaI on the corollary quests for the candidate particle
over the seven annual cycles are calculated taking into
account the time and energy behaviours of the {\it single-hit} experimental data.
In particular, the likelihood function 
requires the agreement: i) of the expectations for the modulated part of the signal 
with the measured modulated behaviour for each detector and for each energy bin; ii) 
of the expectations for the unmodulated component of the signal with the respect 
to the measured differential energy distribution and 
- since ref. \cite{Mod3} - also with the 
bound on recoils obtained by pulse shape discrimination 
from the devoted DAMA/NaI-0 data\cite{Psd96}.
The latter one acts in the likelihood procedure as an experimental upper bound on the unmodulated component of the 
signal and -- as a matter of fact -- as an experimental lower bound on the estimate of the background levels by the maximum
likelihood procedure.
Thus, the C.L.'s, we quote for  
allowed regions, already account for compatibility with the measured differential
energy spectrum and with the measured upper bound on recoils.
In particular, in the following for simplicity, the results of these corollary quests for the
candidate particle are presented in terms of allowed regions
obtained as superposition of the configurations corresponding
to likelihood function values {\it distant} more than $4\sigma$ from
the null hypothesis (absence of modulation) in each of the several 
(but still a limited number) of the possible 
model frameworks considered here. 
Obviously, these results are not exhaustive of the many scenarios
possible at present level of knowledge (e.g. for some other recent ideas 
see \cite{Fre04,Kam03,Sikivie}) and 
larger  
sensitivities than those reported in the following would be reached when including
the effect of other
existing 
uncertainties on assumptions and related parameters \cite{RNC}.

In the most general scenario -- to which the DAMA/NaI target 
nuclei are fully sensitive -- both the SI and the SD components of the cross section 
\begin{figure}[!t]
\begin{center}
\includegraphics[height=9cm]{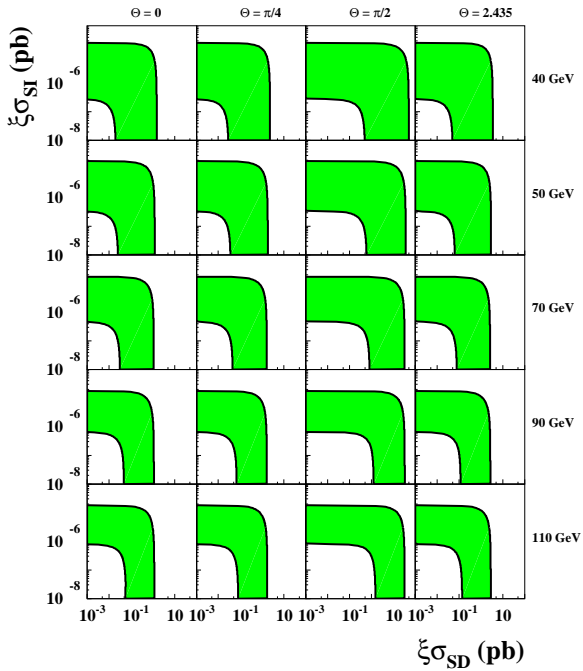}
\end{center}  
\vspace{-0.6cm}
\caption{{\it Case of a WIMP with mixed SI\&SD interaction in the given model
frameworks}. Coloured areas: example of slices (of
the 4-dimensional allowed volume) in the plane 
$\xi \sigma_{SI}$ vs
$\xi \sigma_{SD}$ for some of the possible $m_W$ and $\theta$ values.
Four SD couplings are reported as examples: i)  $\theta$ = 0 ($a_n$
=0 and $a_p \ne$ 0 or  $|a_p| >> |a_n|$) corresponding to a particle
with null SD coupling to neutron; ii) $\theta = \pi/4$ ($a_p = a_n$)
corresponding to a particle with the same SD coupling to neutron and
proton; iii)  $\theta$ = $\pi/2$ ($a_n \ne$ 0 and $a_p$ = 0 
or  $|a_n| >> |a_p|$) corresponding to a particle with null SD
couplings to proton; iv) $\theta$ = 2.435 rad ($ \frac {a_n} {a_p}$
= -0.85) corresponding to a particle with SD coupling through $Z_0$
exchange. The case $a_p = - a_n$ is nearly similar to the case iv).  
Inclusion of other existing uncertainties on parameters and models 
would further extend the regions; for example,
the use of more favourable form factors and/or of more favourable
spin factors than the considered ones 
would move them towards lower cross sections.}
\label{fg:pan_sisd}
\end{figure}
are present. In this general scenario the data give an allowed volume in 
the 4-dimensional space ($m_W$, $\xi \sigma_{SI}$, 
$\xi \sigma_{SD}$, $\theta$). 
Fig. \ref{fg:pan_sisd} just shows slices of this 4-dimensional allowed volume 
in the plane 
$\xi \sigma_{SI}$ vs $\xi \sigma_{SD}$ for some of the possible 
$\theta$ and $m_W$ values for the considered model frameworks.
We just note that experiments 
using either even-spin target nuclei (as Ar and most of Ge, Xe, Te isotopes)
or odd-spin Ge, Xe or Te isotopes
cannot explore most of the allowed volume.
From the given figures it is clear that at present either a purely SI or a purely 
SD or mixed SI\&SD configurations
are compatible with the experimental data of the seven annual cycles. 

\vspace{0.4cm}

\begin{figure}[!ht]
\vspace{-0.6cm}
\begin{center}
\includegraphics[height=4.5cm]{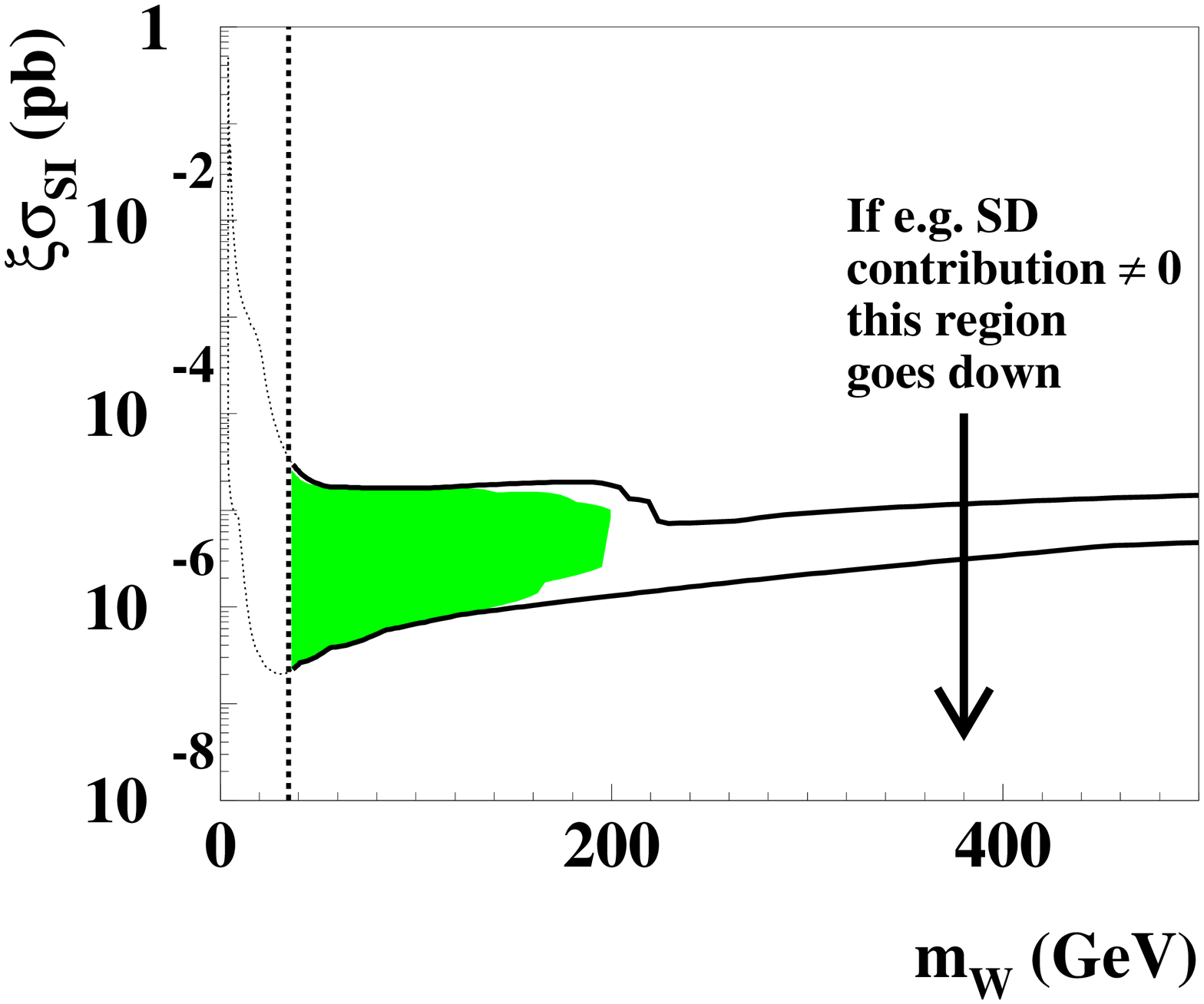}
\includegraphics[height=4.5cm]{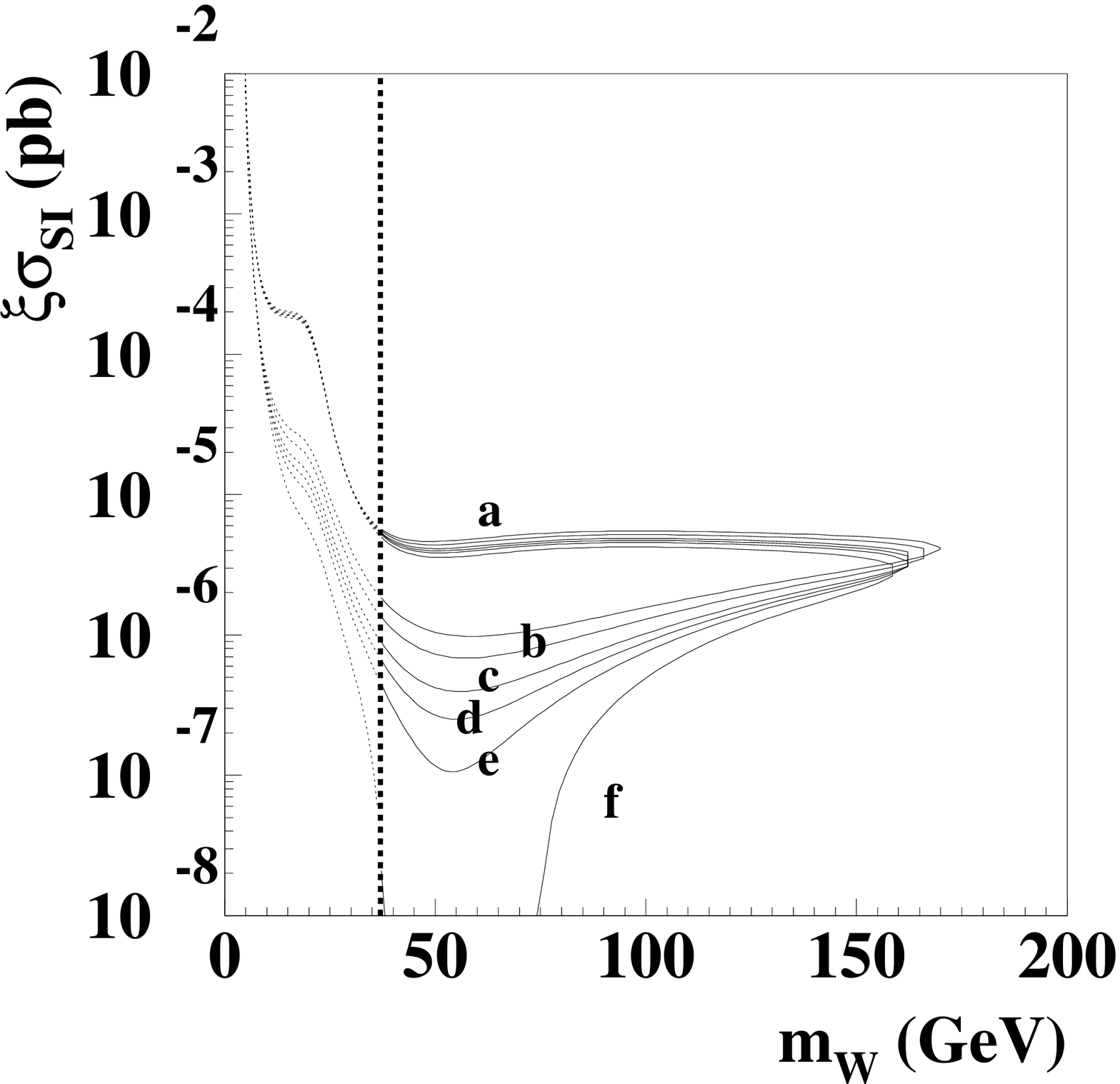}
\end{center}  
\vspace{-0.6cm}
\caption{{\it On the left : Case of a WIMP with dominant SI interaction for the given model
frameworks.} Region allowed in the plane ($m_W$, $\xi
\sigma_{SI}$). The vertical dotted line represents a bound in case
of a neutralino candidate when 
supersymmetric schemes based on GUT assumptions
are adopted to analyse the LEP data; 
the low mass region is allowed for neutralino when other schemes are 
considered and for every other WIMP candidate; see text.
While
the area at WIMP masses above
200 GeV is allowed only for few configurations, the lower one is allowed by most configurations 
(the colored region gathers only those above 
the vertical line) \cite{RNC}.
The inclusion of other existing uncertainties on parameters and models 
would further extend the region; for example, 
the use of more favourable SI form factor for Iodine
alone would move it towards lower cross sections.
{\it On the right :
Example of the effect induced by the inclusion of a SD component
different from zero on  
allowed regions given in the plane $\xi\sigma_{SI}$ vs $m_W$.}
In this example the
Evans' logarithmic axisymmetric $C2$ halo model with
$v_0 = 170$ km/s, $\rho_0$ equal to the maximum value for this model 
and a given set of the parameters' values (see \cite{RNC}) have been considered.
The different regions refer to different SD contributions for the particular case of 
$\theta = 0$:
$\sigma_{SD}=$ 0 pb (a), 0.02 pb (b), 0.04 pb (c), 0.05 pb (d),
0.06 pb (e), 0.08 pb (f). Analogous situation is found for the other 
model frameworks.}
\label{fg:fig_si}
\end{figure}

Often the purely SI interaction
with ordinary matter is assumed to be dominant since e.g.
most of the used target-nuclei are practically
not sensitive to SD interactions (instead, 
$^{23}$Na and $^{127}$I nuclei are fully sensitive) and the theoretical calculations 
and comparisons are even much more complex and uncertain.
Therefore, following the analogous procedure as the general case, we have 
exploited for the same model
frameworks the purely SI scenario, obtaining the allowed region in the plane  
$m_W$ and $\xi \sigma_{SI}$ shown in Fig. \ref{fg:fig_si} -- {\em left}. 
Of course, best fit values of cross section and WIMP mass 
span over a large range in the considered model frameworks.

\begin{figure}[!ht]
\vspace{-0.4cm}
\begin{center}
\leavevmode
\epsfxsize=2.1in \epsfbox{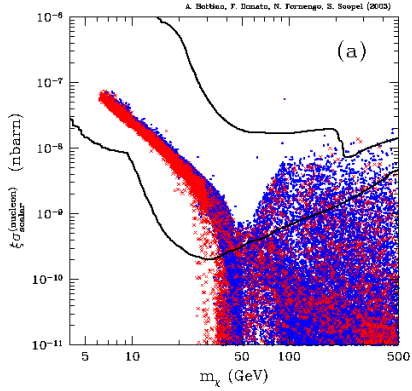}
\end{center}  
\vspace{-0.4cm}
\caption{Figure taken from ref. \cite{Bo04}: theoretical 
expectations of $\xi \sigma_{SI}$ versus $m_W$
in the purely SI coupling for the particular case of a neutralino candidate
in MSSM with gaugino mass unification at GUT scale released; the curve 
surrounds the DAMA/NaI purely SI 
allowed region as
in Fig. \ref{fg:fig_si}-{\it left}.}
\label{figtheo}
\end{figure}

Moreover, configurations with $\xi \sigma_{SI}$ even much lower than those shown in
Fig. \ref{fg:fig_si} -- {\em left} are accessible in case an even small SD contribution 
is present in the interaction as shown as in an example in Fig. \ref{fg:fig_si} -- {\em right}.
Analogous situation is found for other model frameworks.
A comparison of the DAMA/NaI purely SI allowed region (given in Fig. \ref{fg:fig_si} -- {\em left})
with the theoretical expectations for a purely SI coupled neutralino
candidate in a MSSM with gaugino mass unification at GUT scale released is 
shown in Fig. \ref{figtheo} as taken from ref \cite{Bo04}.

\begin{figure}[!ht]
\begin{center}
\vspace{-0.6cm}
\includegraphics[height=4.cm]{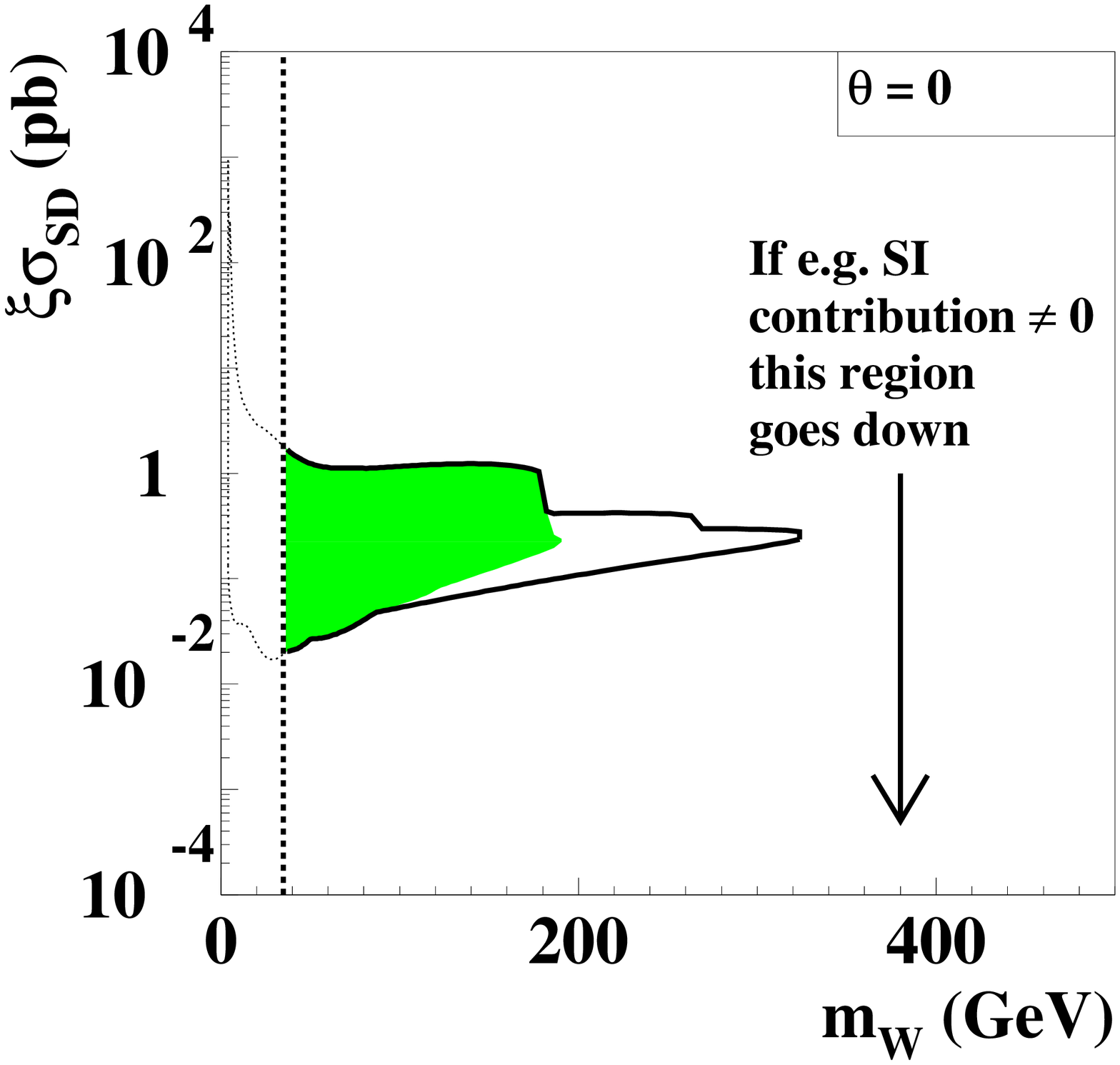}
\includegraphics[height=4.cm]{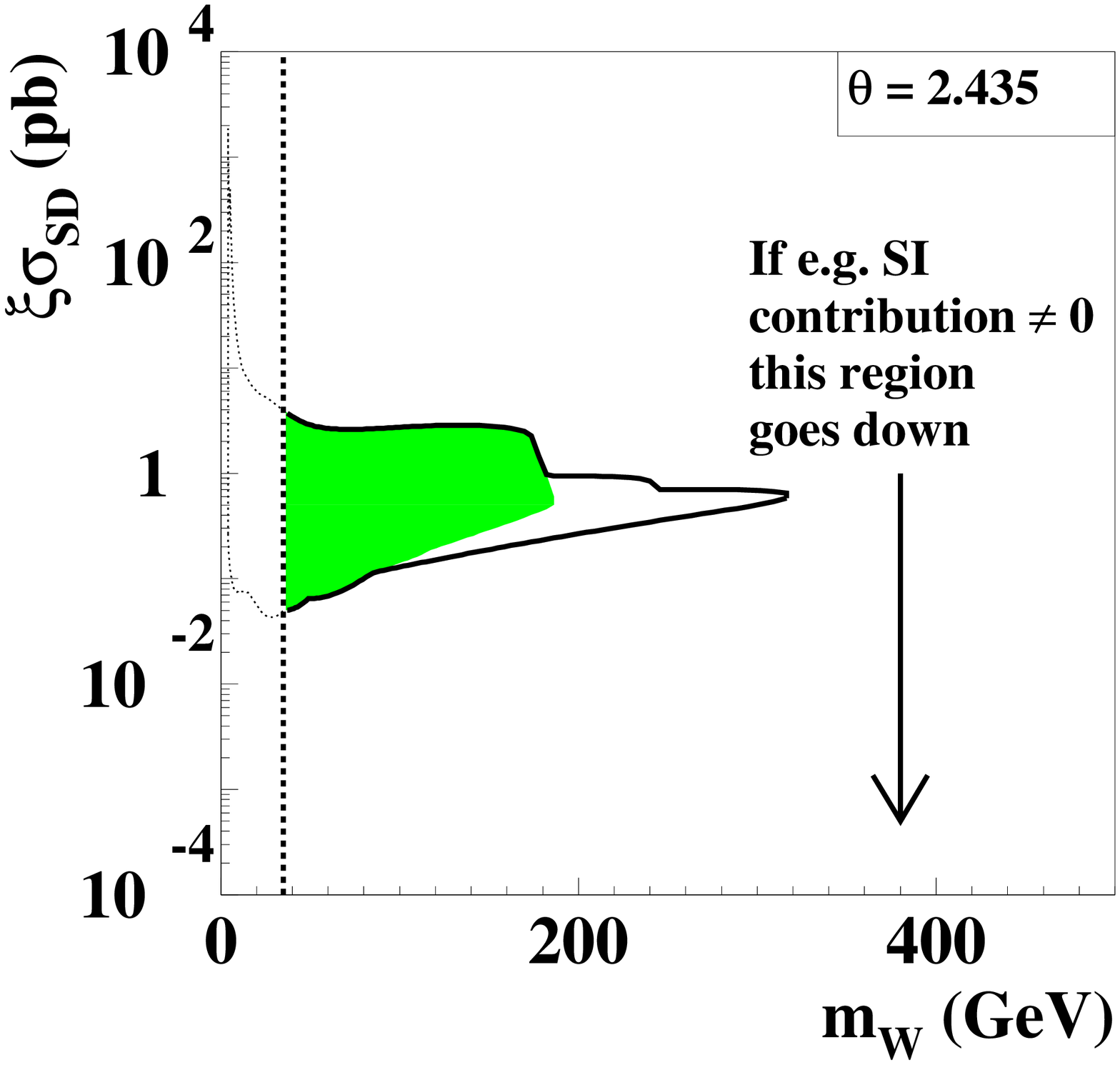}
\end{center}  
\vspace{-0.6cm}
\caption{{\it Case of a WIMP with dominant SD interaction in the given model
frameworks.} Examples of slices (of the 3-dimensional allowed volume) 
in the plane ($m_W$, $\xi \sigma_{SD}$) with $\theta = 0$ and 
$\theta = 2.435$ ($Z_0$ coupling). For the definition of the vertical line and of the coloured
area see previous figure caption.
Inclusion of other existing uncertainties on
parameters and models 
would further extend the SD allowed regions. For example, 
the use of more favourable SD form factors 
and/or more favourable spin factors
would move them towards lower cross sections.
Values of $\xi \sigma_{SD}$ lower than those corresponding to these allowed
regions are possible also e.g. in case of an even small
SI contribution.}
\label{fg:fig_puresd}
\end{figure}

\vspace{0.4cm}

Analogously, one can consider the pure SD coupling.
In this scenario one obtain an allowed volume in the 
3-dimensional space ($m_W$, $\xi \sigma_{SD}$, $\theta$). 
Just examples of some slices of this allowed volume at given $\theta$
is shown in Fig. \ref{fg:fig_puresd}.
Considerations similar to the first case hold.

\vspace{0.4cm}  

\begin{figure}[!htb]
\begin{center}
\vspace{-0.3cm}
\includegraphics[height=7cm]{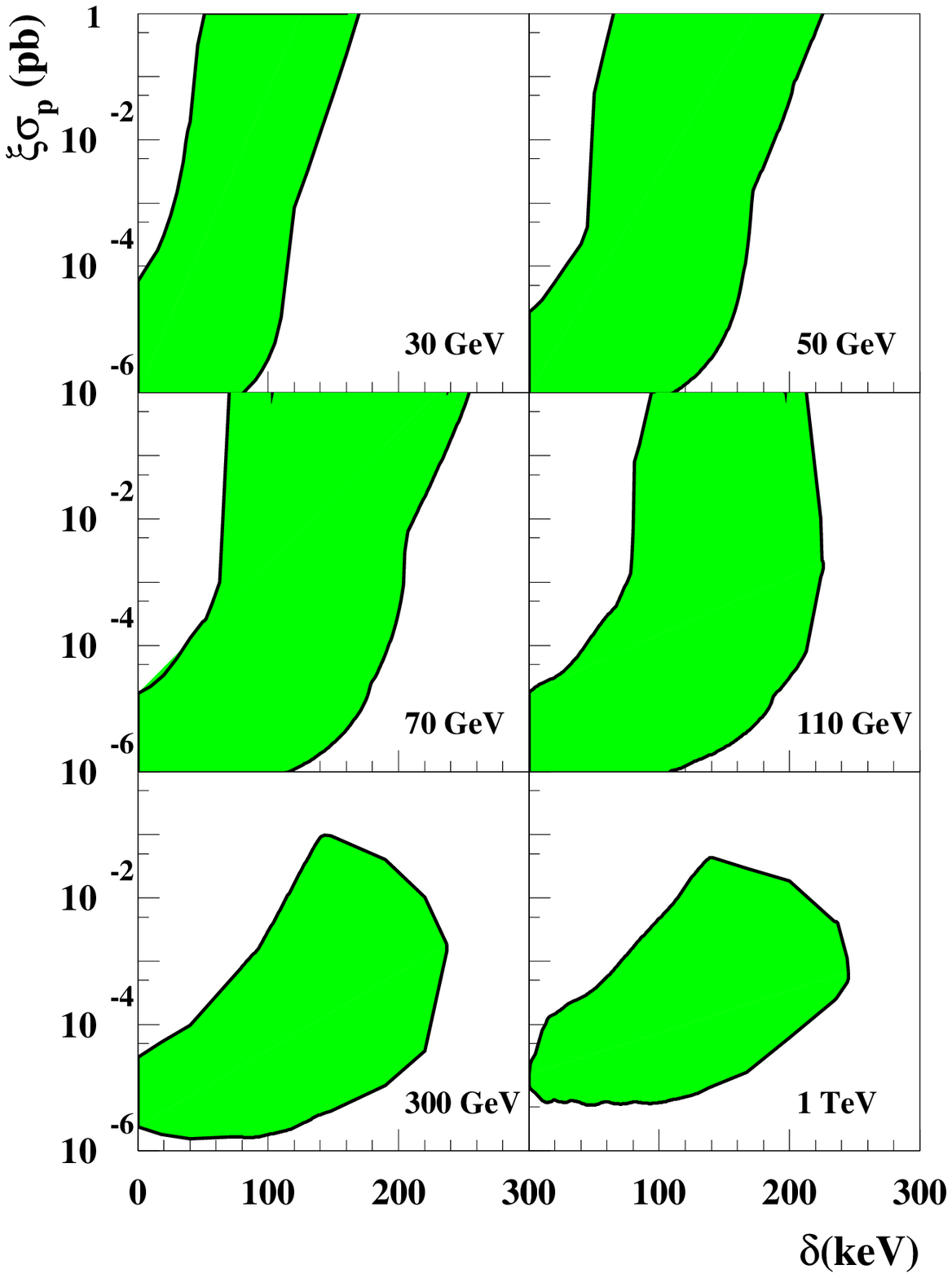}
\end{center}  
\vspace{-0.6cm}
\caption{{\it Case of a WIMP with {\em preferred inelastic} interaction
in the given model frameworks.} Examples of
slices (coloured areas) of the allowed volumes
($\xi \sigma_p$, $\delta$, $m_W$) for some $m_W$ values
for WIMP with {\em preferred inelastic} interaction. 
Inclusion of other existing uncertainties 
on parameters and models 
would further extend the regions; for example,
the use of a more favourable SI form factor for Iodine 
and different escape velocity would move them towards lower cross sections
\cite{RNC}.}
\label{fg:fig_inel}
\vspace{-0.4cm}
\end{figure}

Finally, also the inelastic Dark Matter particle scenario has been analised obtaining 
an allowed volume  in the 3-dimensional space ($\xi \sigma_p$, $m_W$, $\delta$). 
For simplicity, Fig. \ref{fg:fig_inel} shows just few slices of such an allowed 
volume at some given WIMP masses. 
We remind that in these calculations
$v_{esc}$
has been assumed at fixed value
(as in the previous cases), while its present uncertainties can play a significant 
role in this scenario of WIMP with {\em preferred inelastic} scattering.

\vspace{0.2cm}

In conclusion, the data of the seven annual cycles have also 
allowed corollary investigations 
on the features of the Dark Matter particle candidate
in some of the many 
possible scenarios.
However, although several scenarios have been investigated,
these corollary analyses are not exhaustive at all because of the present poor knowledge
on many astrophysical, nuclear and particle Physics needed assumptions.
Thus, further efforts on the topics are in progress and a relevant contribution is
expected by the new DAMA/LIBRA set-up, now running.

\section{The present situation in the field}

\subsection{Direct detection}

As already mentioned, no other experiment, whose result can be directly compared in a model independent 
way with that of DAMA/NaI, is available so far in the field of Dark Matter direct detection.

In fact, most of the activities, started in the 90's, are still at R\&D stage and/or
have released marginal exposures with the respect to the many years of existence and 
to the several used detectors. This is the case of 
CDMS and EDELWEISS experiments, while the Zeplin experiment is more recent \cite{CDMS,Edel,UKXe}.
Since these experiments have claimed to have "excluded" DAMA/NaI, we
will briefly point to the attention of the reader only few arguments.
In particular,  Table \ref{tb:modepcomp1} summarizes some items for comparison. 

\begin{table}[!ht]
\caption{Features of the DAMA/NaI results on the WIMP annual modulation signature over the seven annual cycles 
\cite{RNC}
with those of refs. \cite{CDMS,Edel,UKXe}. }
\vspace{-0.4cm}
\begin{center}
\scriptsize
\begin{tabular}{|c|c|c|c|c|}
\hline \hline
  &  &  & & \\
  & DAMA/NaI  & CDMS-II & Edelweiss-I & Zeplin-I \\
  &  &  & & \\
\hline\hline
  &  &  & &\\
Signature    & Annual  & None  & None & None   \\
  & modulation &  & & \\
\hline
  &  &  &   & \\
Target-nuclei & $^{23}$Na, $^{127}$I & $^{nat}$Ge &  $^{nat}$Ge &  $^{nat}$Xe \\
\hline
  &  &  &   & \\
Technique    & well known &  poorly & poorly & critical optical  \\
  &  & experienced  & experienced & liquid/gas interface \\
  &  &  & &  in this realization\\
\hline
  &  &  & & \\
Target mass &  $\simeq 100$ kg & 0.75 kg & 0.32 kg & $\simeq 3$ kg\\
\hline
  &  &  & & \\
Exposure     &  $\simeq (1.1 \cdot 10^5)$ kg $\cdot$ day & 19.4 kg $\cdot$ day  
                  & 30.5 kg $\cdot$ day  & 280 kg $\cdot$ day\\
\hline
  &  &  & & \\
Depth of the  & 1400 m  & 780 m  & 1700 m & 1100 m \\
experimental site &  &  & & \\
\hline
  &  &  & & \\
Software energy & 2 keV e.e.  & 10 keV e.e.  & 20 keV e.e. & 2 keV e.e.  \\ 
threshold & (5.5 -- 7.5 p.e./keV) &  & & (but: $\sigma/E =100$\%   \\
 &  &  & & mostly   \\
 &  &  & & 1 p.e./keV; \cite{UKXe})\\
 &  &  & & (2.5 p.e./keV \\
 &  &  & & for 16 days; \cite{ZepI}) \\
\hline
  &  &  & & \\
Quenching       & Measured & Assumed  = 1 & Assumed = 1 & Measured  \\
factor          &          &              & (see also \cite{nimqf}) & \\
\hline
  &  &  & & \\
Measured event & $\simeq 1$ cpd/kg/keV & ??, claimed $\gamma$'s        & $\simeq 10^4$ events & $\simeq 100$ cpd/kg/keV \\
rate in low    &                       & larger than CDMS-I            & total                &                         \\
energy range   &                       & ($\simeq 60$ cpd/kg/keV,      &                      &                         \\
               &                       & 10$^5$ events)                &                      &                         \\
\hline
  &  &  & & \\
Claimed events  & & either 0 or 1 & 2 (claimed taken    &  $\simeq$ 20-50 cpd/kg/keV \\
after rejection & &               & in a noisy period!) &  after rejection and \\
procedures      & &               &                     &  ?? after standard PSD \\
                & &               &                     &  \cite{UKXe,ZepI} \\
\hline
  &  &  & & \\
Events satisfying & modulation             & & & \\ 
the signature     & amplitude              & & & \\ 
in DAMA/NaI       & integrated over the    & insensitive & insensitive & insensitive \\ 
                  & given exposure         & & & \\ 
                  & $\simeq 10^3$ events  & & & \\ 
\hline
                & & from few down        & from few down        & \\ 
Expected number & & to zero depending    & to zero depending    & depends on   \\
of events from  & & on the models        & on the models        & the models \\ 
DAMA/NaI effect & & (and on quenching    & (and on quenching    &(even zero) \\
                & & factor)              & factor)              & \\
\hline\hline
\end{tabular}
\end{center}
\label{tb:modepcomp1}
\end{table}

\normalsize

Firstly, let us preliminarily assume as fully correct the "selected" number of events, the energy threshold, the energy scale, etc. 
quoted by those experiments (see Table \ref{tb:modepcomp1}) and let us consider if -- at least under this hypothesis --
their claims might be justified. The answer is obviously not; in fact: i) they give a single model dependent result 
using $^{nat}$Ge or $^{nat}$Xe target, while DAMA/NaI gives a model independent result using $^{23}$Na and $^{127}$I targets;
ii) in the single (of the many possible) model scenario, they consider, they "fix" all the astrophysical, nuclear and particle physics
assumptions at a single choice; the same is even for the experimental and theoretical parameters
values needed in the calculations.
In addition, DAMA/NaI is generally quoted there in an uncorrect, partial and unupdated way
and existing scenarios to which DAMA/NaI is fully sensitive -- on the contrary of the others --
are ignored.

\vspace{0.2cm}

Let us now briefly comment also some of the experimental aspects. 
In particular, the counting rate of the 
Ge bolometers  experiments
is very high and few/zero events are claimed after applying
several strong and hardly safe rejection procedures 
(involving several orders of magnitude).
They usually claim to have an "event by event" discrimination between
{\it noise + electromagnetic background} and {\it recoil + recoil-like (neutrons, end-range alphas, fission fragments,...)} 
events by comparing the bolometer and the ionizing signals for each event, 
but their results are, actually, largely based on "a priori" huge data selections and on the application of other 
preliminar rejection procedures (such as e.g. the one on the so-called surface electrons), which are generally poorly described and
often not completely quantified. 
Moreover, most efficiencies and physical quantities entering in the interpretation of
the claimed selected events have never been properly accounted;
as an example, we mention the case of the bolometer quenching factor of the recoil target nuclei.
In fact, for the bolometer signals the quenching factor (on which the energy threshold and the energy scale rely 
and, hence, also the claimed sensitivity for the given model dependent exclusion plots) 
is arbitrarily assumed to be exactly equal to one. Up to now, only one measurement 
has been made available for a given detector \cite{nimqf}; it offers the value: $0.87 \pm 10\% (stat.) \pm 10\% (syst.) $,
which is --  within the error -- compatible with one, but -- at the same time -- also
compatible with much smaller values. 
Thus, any bolometer result, obtained  without considering e.g. the uncertainties about the unknown value of the quenching factor
and, hence, about the energy threshold and energy scale, has to be considered partial and arbitrary.
For completeness we also mention that the reproducibility of the results over different
running periods has not been proved as well as the values of the effective sensitive
volumes for the read-outs of the two signals for each event and
related quantities; obviously, further uncertainties are present when, as done in
some cases, a neutron background modeling and 
subtraction is pursued in addition.

As regards Zeplin-I \cite{UKXe,ZepI},
a very low energy threshold is claimed (2 keV), although the 
light response is  very poor: between $\simeq$ 1 ph.e./keV \cite{UKXe}
(for most of the time)
and $\simeq$ 2.5 ph.e./keV (claimed for 16 days) \cite{ZepI} \footnote{ 
For comparison we remind that the data of the DAMA/LXe
set-up, which has a similar light response, are analysed by using the much more realistic
and safer software energy threshold of 13 keV \cite{Xe98}.}.
Moreover, a strong data filtering is applied to
the high level of measured counting rate (see Table \ref{tb:modepcomp1})
by hardware vetoes, by fiducial volume cuts and,
largely, by applying down to few keV a standard pulse shape discrimination procedure,
although the LXe scintillation pulse
profiles (pulse decay time $<$ 30 ns) 
are quite similar to the PMT noise events in the lower energy bins and in spite of the
poor light response. Quantitative information on experimental quantities related to the used
procedures has not yet been given \cite{UKXe,ZepI}.

In conclusion, those claims for contradiction have intrinsecally no scientific bases.

\subsection{Indirect detection}

The Dark Matter particles, via their annihilation either in the celestial bodies (such as Earth and Sun) 
or in the Galactic halo could give rise to high energy neutrinos, positrons, antiprotons and gamma rays.
Therefore, they could be {\it indirectly} detected by 
looking either for "upgoing" muons -- produced by $\nu_\mu$ -- in underground, underwater or under-ice detectors
or for antimatter and gamma rays in the space. 
However, it is worth to remark that no direct model independent comparison
can be performed between the results obtained in direct 
and indirect searches.

In the case of "upgoing" muons in terrestrial detectors, 
the expected $\mu$ flux is the key quantity.
However, several sources of uncertainties are present
in the related estimates (and, therefore, in the obtained results) such as e.g.
the assumption that a "steady state" has been reached in the considered celestial body
and the estimate and subtraction
of the existing competing processes, offered by the atmospheric neutrinos.
Model dependent analyses with a similar approach have been carried out 
by large experiments deep underground such as e.g. MACRO and Superkamiokande.
In particular, the case of the neutralino candidate in MSSM has been discussed in 
\cite{Botdm}, showing that their model dependent results were not in conflict
with DAMA/NaI. 

As we mentioned, the annihilation of the Dark Matter 
particles in the galactic halo could also produce antimatter particles and gamma rays. 
The antimatter searches have to be carried out outside the 
atmosphere, i.e. on balloons or satellites. In particular, the Dark Matter particles
annihilation would result in an excess of antiprotons 
or of positrons over an estimated background arising from other possible sources. 
The estimate and subtraction
of such a background together with the influence of the Earth and of the galactic magnetic field
on these particles plays a crucial role on the possibility of a reliable extraction of 
a signal. However, 
at present an excess of positrons  with energy $\simeq 5 - 20$ GeV 
has been suggested by \cite{heat} and other experiments. Interpreted in
terms of Dark Matter particles annihilation 
it gives a result not in conflict 
with the effect observed by DAMA/NaI  \cite{heat}.

As regards the possibility to detect $\gamma$'s from Dark Matter particles annihilation
in the galactic halo, experiments in space are planned. 
However, at present it is difficult to estimate their
possibilities considering 
e.g. the background level, the uncertainties in its reliable estimate 
and subtraction as well as the smallness of the expected signal (even more, 
if a subdominant component would be present) when 
properly calculated with rescaling procedure.
However, in ref. \cite{morse,morse2} 
the presence of a $\gamma$ excess from the center of the Galaxy 
in the EGRET data \cite{egret} has already been suggested.
This excess match with a possible Dark Matter particles annihilation in the galactic halo \cite{morse,morse2} and 
is not in 
conflict with the DAMA/NaI model independent result previously published.

For completeness, we remind that recently it has been suggested \cite{khlopov} that
the positive hints from the indirect detection -- namely the excess of positrons  
and of gamma rays in the space -- and the effect observed by DAMA/NaI
can also be described in a scenario with multi-component Dark Matter in the galactic halo, made of
a subdominant component of heavy neutrinos of the 4$^{th}$ family
and of a sterile dominant component.
\begin{figure}[!ht]
\vspace{-0.4cm}
\begin{center}  
\leavevmode  
\epsfxsize=2.3in \epsfbox{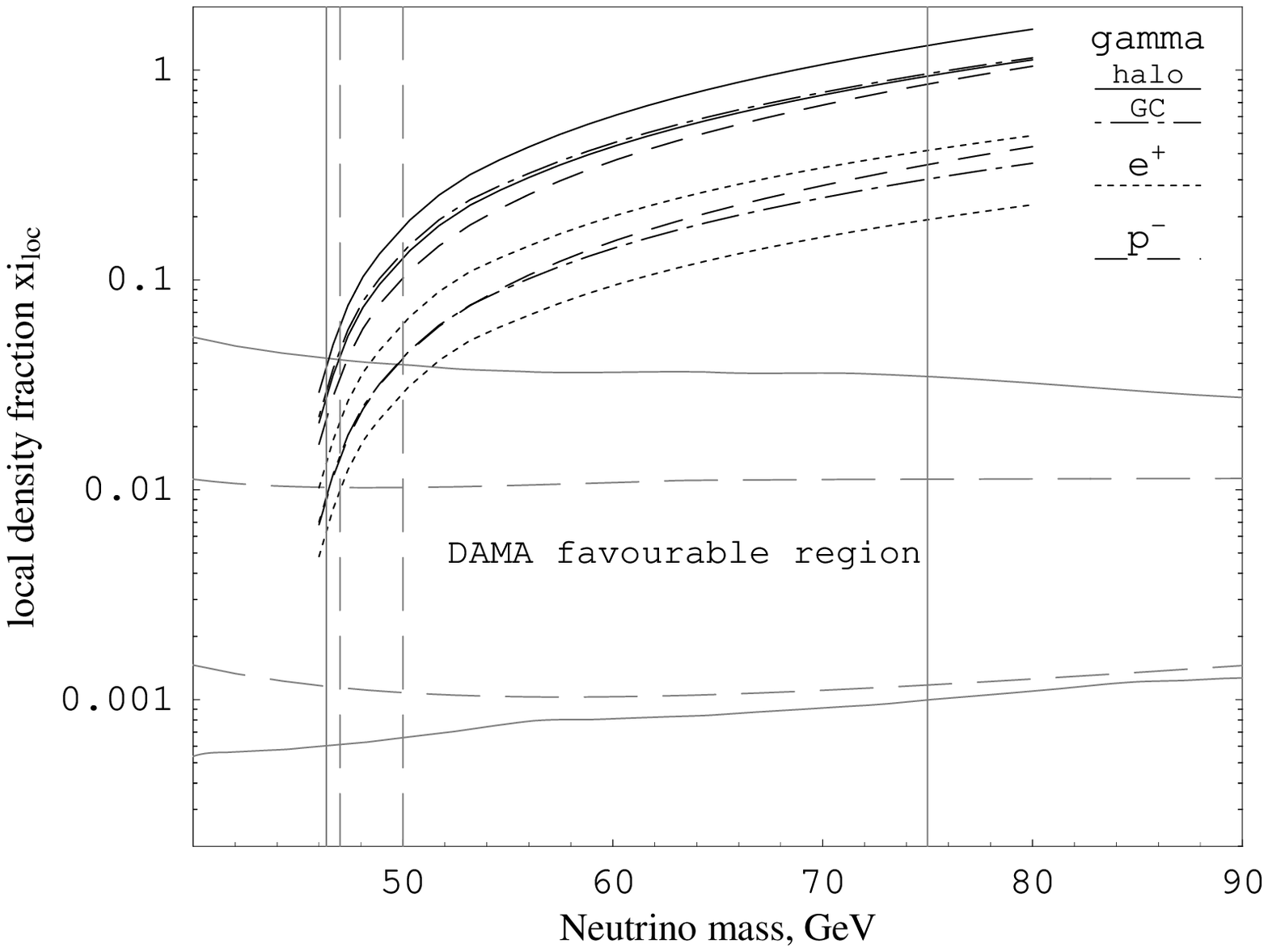}
\epsfxsize=2.3in \epsfbox{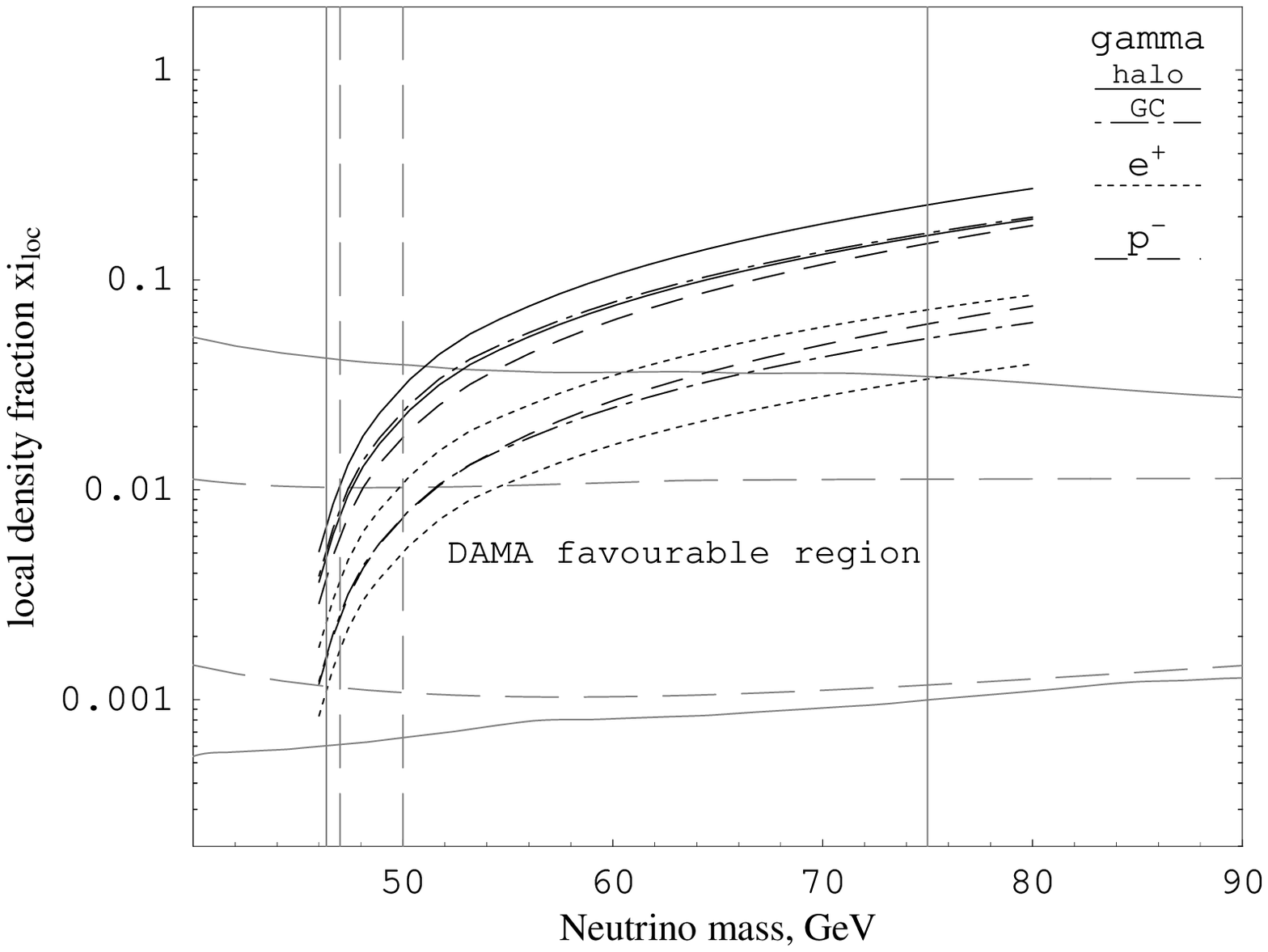}
\end{center}
\vspace{-0.4cm}  
\caption{Figures taken from ref. \cite{khlopov}:
Case of a subdominant heavy 4$^{th}$ neutrino candidate in the plane local density fraction versus
the heavy neutrino mass.
The favorable region obtained from the DAMA/NaI data 
(grey dashed line when using the Evan's halo model; solid line when using the other halo models)
and
the best-fit density parameters deduced from cosmic
gamma-radiation (from halo and galactic center), positron and antiproton
analysis are shown (left panel). 
The effect of the inclusion of possible neutrino clumpiness is also reported (right panel).
See ref. \cite{khlopov} for details.}
\label{figtheo2}
\end{figure}
In particular (see Fig. \ref{figtheo2}), it has been shown that an heavy neutrino 
with mass around 50 GeV can account for all the observations,
while the inclusion of possible clumpiness of
neutrino density as well as new interactions in the heavy neutrino annihilation, etc. can lead to wider mass ranges: 
from about 46 up to about 75 GeV (see ref. \cite{khlopov} for details).

\section{Toward the future: from DAMA/NaI to DAMA/LIBRA and beyond}

The large merits of highly radiopure NaI(Tl) set-up have been demonstrated in the practice by DAMA/NaI
which has been the highest radiopure set-up available in the field, has effectively pursued 
a model independent approach to investigate Dark Matter particles in the galactic halo collecting an 
exposure several orders of magnitude larger than those available in the field and has obtained 
many other complementary or by-products results.

In 1996 DAMA proposed to realize a ton set-up \cite{IDM96} and 
a new R\&D project for highly radiopure NaI(Tl) detectors was funded at that time and carried out for several years 
in order to realize as an intermediate step the second generation experiment, successor of DAMA/NaI, with an exposed 
mass of about 250 kg.

Thus, new powders and other materials have been selected, 
new chemical/physical radiopurification procedures in NaI 
and TlI powders have been exploited, new growing/handling protocols have been developed and
new prototypes have been built and tested.
\begin{figure}[!ht]
\centering
\includegraphics[width=0.4\textwidth]{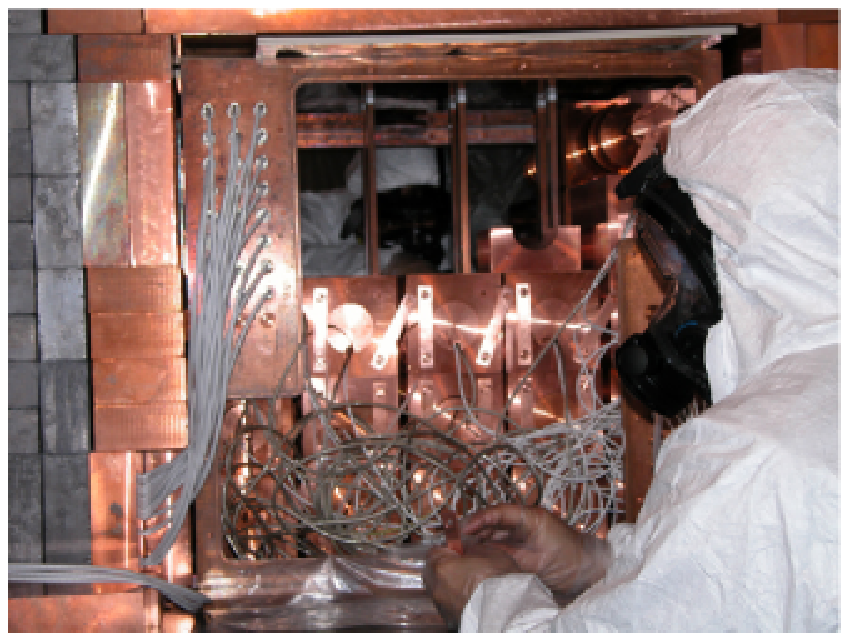}
\includegraphics[width=0.4\textwidth]{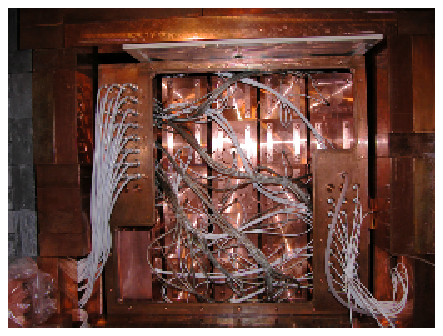}
\includegraphics[width=0.6\textwidth]{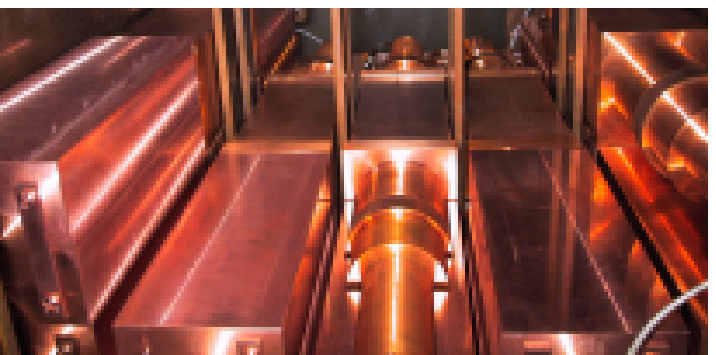}
\caption{{\em Left}: the installation of the 25 NaI(Tl) crystals
(9.70 kg each one) of DAMA/LIBRA in HP Nitrogen atmosphere.
{\em Right}: One of the final stages  of the 
detectors' installation. {\em Bottom}: Picture of the 
photomultiplier shield made of highly radio-pure Copper.
All the used materials have been deeply selected for radiopurity 
(see for example the cables with teflon envelop). All the procedures as well as these photos have been 
carried out in HP Nitrogen atmosphere.}
\label{fg:fig12}
\end{figure}
As a consequence of the results of this second generation R\&D, 
the new experimental set-up DAMA/LIBRA (Large sodium Iodide Bulk for RAre processes),
$\simeq$250 kg highly radiopure NaI(Tl) crystal scintillators  
(matrix of twenty-five $\simeq$ 9.70 kg NaI(Tl) crystals), was funded and realised.
After the completion of the DAMA/NaI data taking in July 2002, the dismounting of DAMA/NaI occurred and the
installation of DAMA/LIBRA started.
In particular, the experimental site as well as many components of the installation itself 
have been implemented (environment, shield of PMTs, wiring, HP Nitrogen system, 
cooling water of air conditioner, electronics and DAQ, etc.). In particular, all 
the Cu parts have been chemically etched before their installation 
following a new devoted protocol and maintained in HP
Nitrogen atmosphere until the installation.
All the procedures performed during the dismounting of DAMA/NaI 
and the installation of DAMA/LIBRA detectors have been carried out in 
HP Nitrogen atmosphere (see Fig.\ref{fg:fig12}).

\begin{figure}[!ht]
\centering
\includegraphics[width=0.5\textwidth]{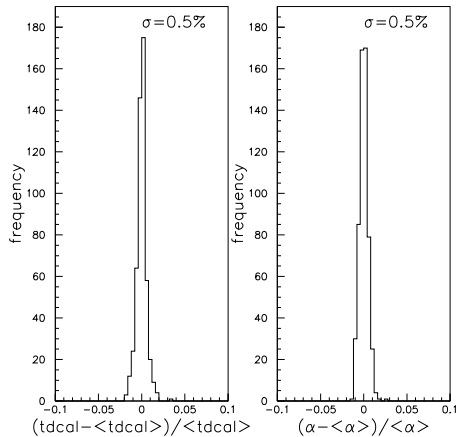}
\vspace{-0.4cm}
\caption{An example of the stability of the calibration factor ({\em tdcal})
and of the ratio of the peaks' positions ($\alpha$) of the 
measured energy distribution of the $^{241}$Am source
during about one year of data taking.}
\label{fg:sta}
\end{figure}

DAMA/LIBRA is taking data since March 2003 and the first data release will, most probably, occur 
when an exposure larger than that of DAMA/NaI will have been collected and analysed in all the aspects. 
Just as an example of the quality of the data taking, Fig. \ref{fg:sta} shows
the stability of the calibration factor and of the ratio of the peaks' positions of the
$^{241}$Am source during about one year of data taking.

The highly radiopure DAMA/LIBRA set-up is a powerful tool for further investigation on the Dark Matter particle component 
in the galactic halo having all the intrinsic merits already mentioned in section 2
and a larger exposed mass, an higher overall radiopurity and improved performances with the respect to DAMA/NaI. 
Thus, DAMA/LIBRA will further investigate the 6.3 $\sigma$ C.L. model independent evidence pointed
out by DAMA/NaI with increased sensitivity in order to reach even higher C.L..
Moreover, it will also offer an increased sensitivity 
to improve corollary quests on the nature of the candidate particle,
trying to disentangle at least among some of the many different possible 
astrophysical, nuclear 
and particle physics models as well as to investigate other new possible scenarios.

In the following some of the main topics -- not yet well known at present and 
which can affect whatever model dependent result and comparison -- 
will be mentioned.
They will be addressed by the highly radiopure DAMA/LIBRA in a way often 
unique and always reliable, cheap and competitive.  
They are:

\begin{itemize} 

\item {\it the velocity and spatial distribution of the Dark Matter particles in the galactic halo}. It has been shown that the
naive description of the galactic halo as an isothermal halo is an unphysical and non-realistic approximation which significantly 
affects model dependent evaluations (exclusion plots, allowed regions, etc.) and comparisons. Other modelings
(not exhaustive at all), many of them based on N-bodies simulations, have been considered in literature and some of them have been
discussed at some extent in \cite{Hep,RNC} and references therein. Some of these models could be significantly discriminated by
DAMA/LIBRA.

\item {\it the effects induced on the Dark Matter particles distribution in the galactic halo by contributions 
from satellite galaxies tidal streams}. Recently it has been pointed out \cite{Fre04} 
that contributions to the Dark Matter particles  
in the galactic halo should be expected from tidal streams from the Sagittarius Dwarf elliptical galaxy. Considering that this
galaxy was undiscovered until 1994 and considering galaxy formation theories, one has to expect that also other satellite galaxies
do exist and contribute as well. In particular, the Canis Major satellite Galaxy has been pointed out as reported in
2003 in ref. \cite{Canis}; it can, in principle, play a very significant role being close to our galactic plane. 
At present, the best way to investigate the presence of a 
stream contribution is to determine in accurate way the phase of the annual modulation, $t_0$, 
as a function of the energy; in fact, for a given halo model $t_0$ would be expected  
to be (slightly) different from 152.5 d and to vary with energy (see Fig. \ref{fg:stream}).

\begin{figure}[!ht]
\centering
\includegraphics[width=250pt]{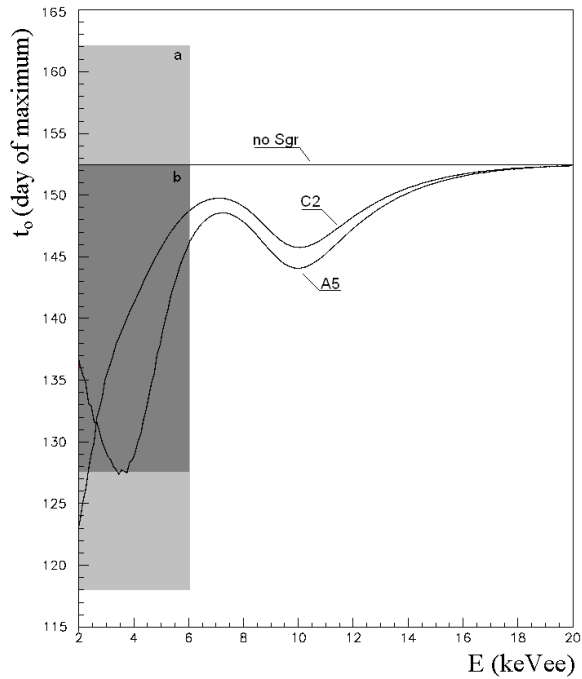}
\vspace{-1.0cm}
\caption{Expected behaviours of the phase, $t_0$, of the annual modulation signal as function of the energy
when considering: i) only galactic halo (``no Sgr''); ii) 
galactic halo ($C2$ halo model with
$v_0 = 220$ km/s, $\rho_0$ equal to the maximum value for this model) 
and a contribution from Sagittarius Dwarf galaxy (``C2''); 
iii) galactic halo ($A5$ halo model with
$v_0 = 220$ km/s, $\rho_0$ equal to the maximum value for this model) 
and a contribution from Sagittarius Dwarf galaxy (``A5''). 
The contributions from Sagittarius Dwarf galaxy 
have been taken in both cases with a density equal to 4\% of $\rho_0$.
The light shadow region is the final result of DAMA/NaI
on the $t_0$ value for the cumulative energy interval (2 -- 6) keV, while the dark shadow region 
is the expectation on $t_0$ assuming an experiment with the
same features as DAMA/NaI, an exposure of $3 \cdot 10^5$ kg $\cdot$ day
and the same central value for $t_0$.}
\label{fg:stream}
\end{figure}   

\item {\it  the effects induced on the Dark Matter particles distribution in the galactic halo by the existence of caustics}.
It has been shown that the continuous infall of Dark Matter particles in the galactic gravitational field can form caustic
surfaces and discrete streams in the Dark Matter particles halo \cite{Sikivie}. The phenomenology to point out a similar scenario 
is analogous to that in the previous item; thus, DAMA/LIBRA can as well test this possibility.

\item {\it the detection of possible "solar wakes"}.
As an additional verification of the possible presence of contributions from streams of Dark Matter particles 
in our galactic halo,
DAMA/LIBRA can investigate also the gravitational focusing effect of the Sun on the Dark Matter particle of a stream.
In fact, one should expect two kinds of enhancements in the Dark Matter particles flow: 
one named "spike", which gives an enhancement of Dark Matter particle density along a line collinear 
with the direction of the incoming stream and of the Sun, and another, named "skirt", which gives
a larger Dark Matter particle density on a surface of cone whose
opening angle depends on the stream velocity. 
Thus, DAMA/LIBRA will investigate such a possibility
with high sensitivity through second-order time-energy correlations.

\item {\it the study of possible structures as clumpiness with small scale size}. Possible structures as clumpiness with small
scale size could, in principle, be pointed out by exploiting a large exposure which can be collected
by DAMA/LIBRA.

\item {\it the coupling(s) of the Dark Matter particle with the $^{23}$Na and $^{127}$I and its nature}.
As mentioned, several large uncertainties are linked to the coupling(s) between the Dark Matter particle and
the target-nuclei. DAMA/LIBRA, exploiting a new large exposure, will allow to better constrain the related aspects.
In addition, analyses in model frameworks suitable for
e.g. mirror Dark Matter \cite{foot} and for particles from multi-dimensional Kaluza-Klein like theories
will be performed as well and compared with the other scenarios analised so far. 

\item {\it scaling laws and cross sections}.
At present just simple scaling laws are used to scale all the nuclear cross sections 
to a common nucleon cross section; however, they are a large source of uncertainties 
in model dependent results and comparisons. For example, recently, it has been pointed out
\cite{Kam03} that, 
even for the neutralino candidate,
these assumptions (which hold in the case of model with one-nucleon current) are arbitrary
when two-nucleon current with pion exchange are introduced. Thus, 
the presence of two target-nuclei in
the NaI(Tl) detectors of DAMA/LIBRA could in principle offer a probe for that in the large exposure 
this experiment will collect.

\end{itemize}

A large work will be faced by DAMA/LIBRA, which is in addition the intrinsecally most sensitive experiment in the field 
of Dark Matter because of its radiopurity, exposed mass and high duty cycle. These qualities will also allow DAMA/LIBRA 
to further investigate with higher sensitivity several other rare processes.

Finally, at present a third generation R\&D effort toward the possible NaI(Tl) ton set-up 
has been funded and related works have already been started.

\section{Conclusion}

DAMA/NaI has been a pioneer experiment running at LNGS for several years and 
investigating as first the model independent WIMP annual modulation signature 
with suitable sensitivity and control of the running parameters. During seven independent 
experiments of one year each one, it has pointed out at 6.3 $\sigma$ C.L. the presence of a modulation
satisfying all the many peculiarities of a WIMP induced effect. Neither systematic effects nor side
reactions able to account for the observed modulation amplitude and to contemporaneously satify 
all the requirements of the signature have been found. 
DAMA/NaI has also pointed out the complexity of corollary investigations on the nature of the candidate  
particle, because of the present poor knowledge on the many astrophysical,
nuclear and particle physics aspects.

A second generation experiment DAMA/LIBRA has been realized and put in operation
since March 2003. This new set-up, having a larger exposed mass 
and an higher overall radiopurity, will offer a significantly increased 
sensitivity to further investigate the Dark Matter particle component in the galactic halo
pointed out at 6.3 $\sigma$ C.L. by DAMA/NaI.
Moreover, further efforts towards the creation of ultimate radiopure NaI(Tl) set-ups are already in progress;
in particular, a third generation R\&D is in progress towards the possible 1 ton set-up we proposed in 1996.

\end{document}